\documentclass[aps]{revtex4}
\usepackage{amsmath}
\usepackage{graphicx}
\usepackage{amsthm}
\newtheorem{theorem}{Theorem}
\newtheorem{lemma}{Lemma}
\newcommand{\vlr}{v_{LR}}

\newcommand{\fs}{\kappa}

\newcommand{\ffn}{{\cal F}}

\newcommand{\be}{\begin{equation}}
\newcommand{\ee}{\end{equation}}
\begin{document}
\title{Making Almost Commuting Matrices Commute}
\author{M.~B.~Hastings}
\affiliation{Microsoft Research, Station Q, Elings Hall, University of California, Santa Barbara, CA 93106}
\begin{abstract}
Suppose two Hermitian matrices $A,B$ almost commute ($\Vert [A,B] \Vert
\leq \delta$).  Are they close to a commuting
pair of Hermitian matrices, $A',B'$, with $\Vert A-A' \Vert,\Vert B-B'\Vert
\leq \epsilon$?  A theorem of H. Lin\cite{hl} shows that this is
uniformly true, in that for every $\epsilon>0$ there exists a $\delta>0$,
independent of the size $N$ of the matrices, for
which almost commuting implies being close to a commuting pair.
However, this theorem
does not specify how $\delta$ depends on $\epsilon$.
We give uniform bounds relating $\delta$ and $\epsilon$.
We provide tighter bounds in the case of
block tridiagonal and tridiagonal matrices and a fully constructive method in that case.
Within the context of quantum measurement, this implies an algorithm
to construct a basis in which we can make a {\it projective} measurement
that approximately measures two approximately commuting operators
simultaneously.  Finally, we comment briefly on the case of approximately
measuring three or more approximately commuting operators using POVMs 
(positive operator-valued measures) instead
of projective measurements.
\end{abstract}
\maketitle

The problem of when two almost commuting matrices are close to matrices
which exactly commute, or, equivalently, when a matrix which is close to normal
is close to a normal matrix, has a long history.  See, for example
\cite{hist1,hist2}, and other references in \cite{hl} where it is
mentioned that the problem dates back to the 1950s or earlier.
Finally in 1995, Lin\cite{hl} proved that
for any $\epsilon>0$, there is a $\delta>0$ such that for all $N$,
for any pair of Hermitian $N$-by-$N$ matrices, $A,B$, with
$\Vert A \Vert, \Vert B \Vert \leq 1$, and $\Vert [A,B] \Vert \leq \delta$,
there exists a pair $A',B'$ with $[A',B']=0$ and $\Vert A-A'\Vert
\leq \epsilon$ and $\Vert B-B' \Vert \leq \epsilon$
This proof was
later shortened and generalized by Friis and Rordam\cite{fr}.  Interestingly,
the same is not true for almost commuting unitary matrices\cite{voiu} or
for almost commuting triplets\cite{voi,krd}.

The importance of the above results is that the bound is uniform in
$N$.  That is, $\delta$ depends only on $\epsilon$.  Unfortunately,
the proofs do not give any bounds on how $\delta$
depends on $\epsilon$.  
In this paper, we present a construction of
matrices $A'$ and $B'$ which enables us to give lower bounds on how small
$\delta$ must be to obtain a given error $\epsilon$.

Specifically, we prove that
\begin{theorem}
\label{mainthm}
Let $A$ and $B$ be Hermitian, $N$-by-$N$ matrices, with
$\Vert A \Vert, \Vert B \Vert \leq 1$.  Suppose
$\Vert [A,B] \Vert \leq \delta$.  Then, there exist Hermitian,
$N$-by-$N$ matrices $A'$ and $B'$ such that
\begin{itemize}
\item[{\bf 1}:] $[A',B']=0$.

\item[{\bf 2}:] $\Vert A'-A \Vert \leq \epsilon(\delta)$ and $\Vert B'-B \Vert \leq
\epsilon(\delta)$,
with
\be
\label{mainthmerbnd}
\epsilon(\delta)=E(1/\delta) \delta^{1/5},
\ee
where
the function $E(x)$ grows slower than any
power of $x$.  The function $E(x)$ does not depend on $N$.
\end{itemize}
\end{theorem}
Throughout this paper, we use $\Vert ... \Vert$
to denote the operator norm of a matrix, and
$|....|$ to denote the $l^2$-norm of a vector.

The proof of theorem (\ref{mainthm}) involves first constructing a related problem
involving a block tridiagonal matrix, $H$, and a block identity matrix $X$ (we use
the term ``block identity matrix" to refer to a block
diagonal matrix that is proportional to the identity matrix in each block).
For such matrices we prove the
theorem
\begin{theorem}
\label{btridthm}
Let $X$ be a block identity Hermitian matrix and 
let $H$ be a block tridiagonal Hermitian matrix, with the $j$-th block of $X$ equal to
$c+j\Delta$ times the identity matrix, for some constants $c$ and
$\Delta$.  Let $\Vert H \Vert, \Vert X \Vert \leq 1$.
Then, there exist Hermitian matrices $A'$ and $B'$ such that
\begin{itemize}
\item[{\bf 1}:] $[A',B']=0$.

\item[{\bf 2}:] $\Vert A'-H \Vert \leq \epsilon'(\Delta)$ and $\Vert B'-X \Vert \leq
\epsilon'(\Delta)$, with
\be
\label{btridthmerbnd}
\epsilon'(\Delta)=E'(1/\Delta) \Delta^{1/4},
\ee
where
the function $E'(x)$ grows slower than any
power of $x$.  The function $E'(x)$ does not depend on the dimension of the matrices.
\end{itemize}
\end{theorem}
After proving these results, we prove a tighter bound in the case where
$H$ is a tridiagonal matrix, rather than a block tridiagonal matrix:
\begin{theorem}
\label{tridtheorem}
Let $X$ be a diagonal Hermitian matrix and 
let $H$ be a tridiagonal Hermitian matrix, with the $j$-th diagonal entry of $X$ equal to
$c+j\Delta$, for some constants $c$ and
$\Delta$.  Let $\Vert H \Vert, \Vert X \Vert \leq 1$.
Then, there exist Hermitian matrices $A'$ and $B'$ such that
\begin{itemize}
\item[{\bf 1}:] $[A',B']=0$.

\item[{\bf 2}:] $\Vert A'-H \Vert \leq \epsilon'(\Delta)$ and $\Vert B'-X \Vert \leq
\epsilon'(\Delta)$, with
\be
\epsilon'(\Delta)=E''(1/\Delta) \Delta^{1/2},
\ee
where
the function $E''(x)$ grows slower than any
power of $x$.  The function $E''(x)$ does not depend on the dimension of the matrices.
\end{itemize}
\end{theorem}

The proofs rely heavily on ideas relating to
Lieb-Robinson bounds\cite{lrbounds1,lrbounds2,lrbounds3,lrbounds4}.  These
bounds, combined with appropriately chosen filter functions,
have been used in recent years in Hamiltonian complexity
to study the dynamics and
ground states of quantum systems, obtaining results such as
a higher dimensional Lieb-Schultz-Mattis theorem\cite{lrbounds2},
a proof of exponential decay of correlations\cite{loc},
studies of dynamics out of equilibrium\cite{quench,tj,bv},
new algorithms for simulation of quantum systems\cite{tjocert,tjo,qbp,tjoa,obs},
an area law for entanglement entropy for general interacting
systems\cite{areal}, study of
harmonic lattice systems\cite{ng},
a Goldstone theorem with fewer assumption\cite{goldstone},
and many others.  The present paper represents a different application,
to the study of almost commuting matrices.  

Before beginning the proof, we give some discussion of physics intuition
behind the result.  The next few paragraphs are purely to
motivate the problems from a physics viewpoint.
In the last section on quantum measurement and in the
discussion at the end we give additional
applications to quantum measurement and construction of Wannier functions.
The section on quantum measurement is intended to be self-contained.
As mentioned, we begin by relating this
problem to the study of block tridiagonal matrices.  We then interpret
the matrix $H$ as a Hamiltonian for a single particle moving
in one dimension,
and apply the Lieb-Robinson bounds.  
The result (\ref{btridthmerbnd})
implies that we can construct a complete orthonormal basis of states
which are simultaneously localized in both position ($X$) and
energy ($H$).  It is certainly easy to construct an overcomplete
basis of states which is localized in both position and energy,
by considering, for example, Gaussian wavepackets.  The interesting
result is the ability to construct an orthonormal basis which satisfies this.

Additional physics intuition can be obtained by
considering the case where
$H$ is a tridiagonal matrix with $0$ on the diagonal and elements just
above and below the diagonal equal to $1$, and where $X$ is a diagonal
matrix with entries $1/N,2/N,...$.  We refer to this as a
uniform chain.  
In the uniform chain case, if we define a new
matrix $H'$ by randomly
perturbing $H$, replacing each diagonal element of $H$ with a small
diagonal number chosen at random, the eigenvectors of $H'$ are
localized with high probability\cite{loceigfn,loceignfn2}.  Then, we can
construct a matrix $X'$ which exactly commutes with $H'$ as
follows:  if
$v$ is an eigenvector of $H'$, we choose it to have eigenvalue for
$X'$ equal to $(v,Xv)$.  Then, since the eigenvectors are localized,
we find that $\Vert X-X' \Vert$ is small.  The difference
$\Vert X -X' \Vert$ depends on the localization length which depends
inversely on the amount of disorder, while the difference
$\Vert H-H' \Vert$ depends on the amount of disorder.  
Unfortunately, we do not
have a good enough understanding of the effect of disorder for
matrices $H$ which are block tridiagonal, rather than just tridiagonal,
to turn this approach into a proof for general $H$ and $X$, and
thus we rely on an alternative, constructive approach.

\section{Proof of Main Theorem}
We now outline the proof of theorem (\ref{mainthm}).  The proof
is described by the following algorithm:
\begin{itemize}
\item[{\bf 1}:] Construct $H$ from $A$ as described in section
(\ref{h0sub}) and lemma
(\ref{h0con}).  We will bound
$\Vert H-A \Vert$.

\item[{\bf 2}:] Construct $X$ from $B$ as described in section
(\ref{xsub}).  We will bound $\Vert X-B \Vert$.  In a basis of
eigenvalues of $X$, the matrix $H$ will be block tridiagonal.

\item[{\bf 3}:]  Construct a new basis
as described in section (\ref{ucon}) such that in this basis
$H$ is close to a block diagonal
matrix.  That is, we will bound the operator norm of the
block off-diagonal part of $H$.   The blocks will be different from the
blocks considered in step {\bf (2)} above and will be larger.
Further, we will show that $X$ is close to a block identity matrix
in this basis.

\item[{\bf 4}:]  
Set $A'$ to be the block diagonal part of $H$ in the basis constructed
in step {\bf (3)} and set $B'$ to the block identity matrix constructed
in step {\bf (3)}, so that $[A',B']=0$.

\end{itemize}
This algorithm involves several choices of constants.  
In a final section,
(\ref{erbnds}), we indicate how to pick the constants to
obtain the error bound (\ref{mainthmerbnd}).  The key step will be step $3$.

\section{Reduction to Block Tridiagonal Problem}

The first
two steps of the proof above
{\bf (1,2)} reduce theorem (\ref{mainthm}) to theorem
(\ref{btridthm}), while the last two steps {\bf (3,4)} prove
theorem (\ref{btridthm}).  In this section we present the
first two steps.

\subsection{Construction of Finite-Range $H$}
\label{h0sub}

We begin
by constructing matrix $H$ as given in the following lemma, where the
constant $\Delta$ will be chosen later.

\begin{lemma}
\label{h0con}
Given Hermitian matrices $A$ and $B$, with $\Vert [A,B]\Vert \leq \delta$,
for any $\Delta$
there exists a Hermitian matrix $H$ with the following properties. 
\begin{itemize}
\item[{\bf 1}:] $\Vert [H,B] \Vert \leq \delta$.  
\item[{\bf 2}:] For any two vectors $v_1,v_2$ which are eigenvectors
of $B$ with corresponding eigenvalues $x_1,x_2$, and with
$|x_1-x_2|\geq \Delta$, we have $(v_1,H v_2)=0$.
\item[{\bf 3}:] $\Vert A-H \Vert \leq \epsilon_1$, with
$\epsilon_1=c_0 \delta/\Delta$, where
$c_0$ is a numeric constant given below.
\end{itemize}
\begin{proof}
We define
\be
\label{hpdef}
H=\Delta \int {\rm d} t \exp(i B t) A \exp(-i B t) f(\Delta t),
\ee
where the function $f(t)$ is defined to have the Fourier transform
\begin{eqnarray}
\label{tfft}
\tilde f(\omega)=(1-\omega^2)^3,\quad |\omega|\leq 1 \\
\nonumber
\tilde f(\omega)=0, \quad  |\omega|\geq 1,
\end{eqnarray}
and hence the Fourier transform of $f(\Delta t)$ is supported on the
interval $[-\Delta,\Delta]$.
Properties (1) and (2) follow immediately from Eq.~(\ref{hpdef}).
Property (3) follows from
\begin{eqnarray}
\label{seclin}
\Vert H-A \Vert 
&=& 
\Vert \Bigl( \Delta \int {\rm d} t \, \exp(i B t) A \exp(-i B t)
f(\Delta t) \Bigr)-A \Vert
\\ \nonumber
&=&
\Vert \Delta \int {\rm d} t \, \Bigl( \exp(i B t) A \exp(-i B t) -A \Bigr) f(\Delta t) \Vert
\\ \nonumber
& \leq & 
\Delta \int {\rm d} t \, \Vert \Bigl( \exp(i B t) A \exp(-i B t) -A \Bigr) \Vert \; | f(\Delta t) |
\\ \nonumber
&\leq & \Delta \int {\rm d}t \, |t|\Vert [A,B] \Vert \; | f(\Delta t) | \\ \nonumber
& \leq & \delta \Delta
\int {\rm d}t \, |t f(\Delta t) | \\ \nonumber
&=&
c_0 \delta/\Delta,
\end{eqnarray}
where
we define the constant $c_0$ by
\be
\label{c0d}
c_0= \int {\rm d} t |t f(t)|.
\ee
The second line in Eq.(~\ref{seclin}) follows because $\tilde f(0)=1$ so that $\Delta \int {\rm d} t A f(\Delta t)=A$.
Note that since the first and second derivatives of $\tilde f(\omega)$ vanish at
$\omega=\pm 1$, the function $f(t)$ decays as $1/t^3$ for large $t$ and
hence $c_0$ is finite.  
Since $\tilde f$ is an even function, $H$ is Hermitian.

Note that the precise form of the function
$f(t)$ is unimportant: all we require is that $\tilde f(0)=1$; that
$\tilde f$ is supported on the interval $[-1,1]$; that $\tilde f$
is sufficiently smooth that $f(t)$ decays fast enough for the integral
over $t$ (\ref{c0d}) to converge; and that $\tilde f$ is an even function.  
\end{proof}
\end{lemma}

{\bf Remark:} In a basis of eigenvectors of $B$,
property (3) in the above lemma implies that $H$ is
``finite-range", in that the off-diagonal elements are vanishing for
sufficiently large $|x_1-x_2|$.  The next theorem is a Lieb-Robinson
bound for such finite range Hamiltonians, similar to those proven
for many-body Hamiltonians\cite{lrbounds1,lrbounds2,lrbounds3,lrbounds4}.
This result is also similar to results on the decay of entries of smooth functions
of matrices proven in \cite{dms,bg}.

We now introduce some terminology.  Given two sets of real
numbers, $S_1,S_2$, we define
\be
{\rm dist}(S_1,S_2)=\min_{x_1 \in S_1,x_2 \in S_2} |x_1-x_2|.
\ee

{\bf Remark:}
The reason for introducing this ``distance function" is that we think of $H$
as defining the Hamiltonian for a one-dimensional, finite- range quantum
system, with different ``sites" of the system corresponding to different
eigenvectors of $B$, and then the distance function is the distance
between different sets of sites.

Further, we
say that a vector $w$ is ``supported on set $S$ for position operator
$B$" if $w$ is a linear combination of eigenvectors of $B$ whose
corresponding eigenvalues are in set $S$.  Finally, for any set
$S$ we define the projector
$P(S,B)$ to be the projector onto eigenvectors of $B$ whose
corresponding eigenvalues lie in set $S$.  We now give the Lieb-Robinson
bound:

\begin{theorem}
Let $H$ have the properties
\begin{itemize}
\item[{\bf 1}:] $\Vert H \Vert \leq 1$.
\item[{\bf 2}:] For any two vectors $v_1,v_2$ which are eigenvectors
of $B$ with corresponding eigenvalues $x_1,x_2$, and with
$|x_1-x_2|\geq \Delta$, we have $(v_1,H v_2)=0$.
\end{itemize}
Define
\be
\vlr=e^2 \Delta.
\ee
Then, for any vector $v$ supported on a set $S_1$ for position operator
$B$, and for any projector $P(S_2,B)$, we have
\be
|P(S_2,B) \exp(-i H t) v| \leq e^{-{\rm dist}(S_1,S_2)/\Delta} |v|
\ee
for any
\be
\label{tbnd}
|t|\leq {\rm dist}(S_1,S_2)/\vlr.
\ee
\begin{proof}
Expand $\exp(-i H t) v$ in a power series as
$v-i H t v-(H^2/2) t^2 v+...$.  Then, by assumption,
$P(S_2,B) (-i t)^n (H^n/n!) v$ vanishes for $n<{\rm dist}(S_1,S_2)/\Delta$.
Let $m= \lceil {\rm dist}(S_1,S_2)/\Delta \rceil$.
Then,
\begin{eqnarray}
|\sum_{n\geq m} (-i t)^n (H^n/n!) v| &\leq &
\sum_{n\geq m} (|t|^n/n!) |v|
\\ \nonumber
& \leq &
\frac{1}{e}
\sum_{n\geq m}
(e|t|/n)^n |v| \\ \nonumber
& \leq &
\frac{1}{e}
(e|t|/m)^m \frac{1}{1-e|t|/m} |v|.
\end{eqnarray}
For the given $\vlr$ and $t$, the result follows.
\end{proof}
\end{theorem}

{\bf Remark:} the proof of this Lieb-Robinson bound is significantly
simpler than the proofs of the corresponding bounds for many-body systems
considered elsewhere.  The power series technique used here does not work
for such systems.

\subsection{Construction of $X$}
\label{xsub}

In this subsection, we construct the matrix $X$ from $B$.
We define a function $Q(x)$ by 
\be
Q(x)=\Delta \lfloor x/\Delta +1/2 \rfloor.
\ee
Then, we set
\be
X=Q(B).
\ee

Note that $|Q(x)-x|\leq \Delta/2$ for all $x$, and $Q(x)/\Delta$ is always
an integer.  Then,
\be
\Vert X-B \Vert \leq \epsilon_2,
\ee
with
\be
\epsilon_2=\Delta/2.
\ee
By ${\bf (2)}$ in lemma (\ref{h0con}), the matrix $H$ is a block
tridiagonal matrix when written in a basis of eigenstates of $X$,
with eigenvalues of $X$ ordered in increasing order.

\section{Construction of New Basis}
\label{ucon}

In this section we construct the basis to make $H$ close to
a block diagonal matrix and $X$ close to a block identity matrix.
This completes step ({\bf 3}) of the construction of $A'$ and $B'$.
We refer to the basis constructed in this step as the ``new basis" and
we refer to the basis in which $X$ is diagonal as the ``old basis".

There will be a total of $n_{cut}+1$ different blocks in the new
basis, where $n_{cut}$
is chosen later.  Before constructing the new basis, we give some definitions.
We define an interval $I_i$ by
$I_i=[-1+2(i-1)/n_{cut},-1+2i/n_{cut})$ for $1\leq i<n_{cut}$ and
$I_i=[-1+2(i-1)/n_{cut},-1+2i/n_{cut}]$ for $i=n_{cut}$.
Let $J_i$ be the 
matrix given by projecting $H$ onto
the subspace of eigenvalues of $X$ lying in this interval $I_i$,
and call this subspace
${\cal B}_i$.  Then, in the old basis of eigenvalues of $X$, $J_i$ is 
block tridiagonal with
at least $L$ different blocks, 
where $L=\lfloor (2/n_{cut}\Delta)-1\rfloor$ (some of these
blocks might have dimension zero if $B$ happens to have fewer than
$L$ distinct eigenvalues in that interval).  We will choose $n_{cut}$ later
so that $L>>1$ and so the new basis will have fewer blocks than the old basis.
Before constructing the new basis we need the following lemma.

We claim that:
\begin{lemma}
\label{mainlemma}
Let $J$ be a Hermitian block tridiagonal matrix, with $\Vert J \Vert \leq 1$,
acting on a space ${\cal B}$.
Let there be $L$ blocks, so that the space ${\cal B}$ has
$L$ orthogonal subspaces, which we write
${\cal V}_j$ for $j=1,...,L$, with
$(v,Jw)=0$ for $v\in {\cal V}_i$ and $w \in {\cal V}_j$ with
$|i-j|>1$.
Then, there exists a space ${\cal W}$ which is a
subspace of ${\cal B}$ with the following properties:
\begin{itemize}
\item[{\bf 1}:]
The projection of any normalized
vector $v \in {\cal V}_1$ onto
the orthogonal complement of ${\cal W}$ has
norm bounded by $\epsilon_3$ where
$\epsilon_3$ is equal to $1/L^{1/3}$ times
a function
growing slower than any power of $L$.

\item[{\bf 2}:]
For any normalized vector
$w\in {\cal W}$, the projection of
$J w$ onto
the orthogonal complement of ${\cal W}$ has
norm bounded by $\epsilon_4$, where
$\epsilon_4$ is equal to $1/L^{1/3}$ times
a function
growing slower than any power of $L$.

\item[{\bf 3}:]
The projection of any normalized
vector $v \in {\cal V}_L$ onto
${\cal W}$ has
norm bounded by $\epsilon_5$, where
$\epsilon_5$ is equal to
a function decaying faster
than any power of $L$.
\end{itemize}
\begin{proof}
This lemma is the key step in the proof of the main theorem, and
the proof of this lemma is given in the next section.
\end{proof}
\end{lemma}

For each $i$, $1\leq i \leq n_{cut}$, we apply lemma (\ref{mainlemma}) to
the matrix $J=J_i$ defined on the space ${\cal B}={\cal B}_i$.
For given $i$, we refer to the space ${\cal W}$ as
constructed in lemma (\ref{mainlemma}) as ${\cal W}_i$ and we refer to
the subspaces ${\cal V}_j$ defined in lemma (\ref{mainlemma})
as ${\cal V}_j(i)$.
Let ${\cal B}_i$ have dimension $D_B(i)$ and let ${\cal W}_i$ have
dimension
$D_W(i)$.  
Let ${\cal W}^{\perp}_i$ denote the $D_B(i)-D_W(i)$-dimensional space which
is the orthogonal complement of ${\cal W}_i$.
Let ${\cal V}_j(i)$ have dimension $d_j(i)$.
By
properties {\bf (1,2)} in lemma (\ref{mainlemma}), $D_B(i) \geq d_1(i)$ and
$D_B(i)\leq D_W(i)-d_L(i)$.

The new basis has $n_{cut}+1$ blocks, which we label by $i=0,1,...,n_{cut}$.
For $1\leq i < n_{cut}$,
we define the $i$-th block of the new basis
to be the space spanned by ${\cal W}_{i+1}$ and ${\cal W}^{\perp}_{i}$.  For
$i=0$, the $i$-th block is the space ${\cal W}_{i+1}={\cal W}_1$.  For $i=n_{cut}$,
the $i$-th block is the space ${\cal W}^{\perp}_i={\cal W}^{\perp}_{n_{cut}}$.

Then, the matrix $H$ is a block band matrix in this new basis.  
The matrix $H$ will have terms on the main diagonal, on the diagonals
directly above and below the main diagonal, and on the diagonals above
and below those, so it only has terms within distance two of the main
diagonal.
The
block-off-diagonal terms above and below the main diagonal
arise from three sources.  First, the matrix
$J_i$ contains non-vanishing matrix elements between the spaces ${\cal W}_i$
and ${\cal W}^{\perp}_i$, and those spaces are now in different blocks.
However, by property ${\bf (2)}$ in lemma (\ref{mainlemma}), these
matrix elements are bounded by $\epsilon_4$.  Second, there are 
non-vanishing matrix elements between the subspace ${\cal W}_{i-1}^{\perp}$
and ${\cal V}^i_1$, and ${\cal V}^i_1$ may not be completely contained
in subspace ${\cal W}_i$.  However, by property {\bf (1)} in lemma
(\ref{mainlemma}), these contribute only $\epsilon_3$ to the norm
of the block-off-diagonal terms of $H$ in the new basis.
Third, there are non-vanishing matrix
elements between ${\cal W}_{i}$ and ${\cal V}^i_L$, and ${\cal V}^i_L$ may not be completely contained
in subspace ${\cal W}^{\perp}_i$.  However, by property {\bf (1)} in lemma
(\ref{mainlemma}), these contribute only $\epsilon_5$ to the norm
of the block-off-diagonal terms of $H$ in the new basis.  Therefore,
these block-off-diagonal terms in $H$ are bounded in operator norm by
\be
\label{sumerr}
2 (\epsilon_3+\epsilon_4+\epsilon_5).
\ee
The terms which are on the diagonals two above or two below the main diagonal
arise from the fact that there are non-vanishing matrix elements between
${\cal V}^i_L$ and ${\cal V}^{i+1}_1$, and ${\cal V}^i_L$ has some
non-vanishing overlap with ${\cal W}_i$ and ${\cal V}^{i+1}_1$ has
some non-vanishing overlap with ${\cal W}^\perp_{i+1}$.  These
terms are bounded by $\epsilon_3\epsilon_5$.

Define 
$B'$ to be the block identity matrix (in the new basis) which is
equal to $-1+2i/n_{cut}$ times the identity matrix in the
$i$-th block.
Since each block $i$ in the new basis lies within the
space spanned by ${\cal B}_i$ and ${\cal B}_{i+1}$ we have
\be
\label{bdiff}
\Vert B'-B \Vert \leq 2/n_{cut}.
\ee

{\bf Remark: }Here is a sketch of the above procedure, in a case where $H$ has $8$ blocks
and $n_{cut}=2$.  The matrix originally looks like
\be
\begin{pmatrix}
... & ... \\
... & ... & ... \\
& ... & ... & ... \\
&& ... & ... & ... \\
&&& ... & ... & ... \\
&&&& ... & ... & ... \\
&&&&& ... & ... & ... \\
&&&&&& ... & ... \\
\end{pmatrix}
\ee
where the $...$ indicate non-vanishing entries.  We combine the entries in the first $4$
blocks into a matrix $J_1$ and the entries in the last $4$ into a matrix $J_2$ so $H$
looks like
\be
\label{hj1j2}
\begin{pmatrix}
J_1 & ... \\
... & J_2,
\end{pmatrix}
\ee
where the $...$ couples only the $L$-th block of space ${\cal B}_1$ to the $1$-st
block of space ${\cal B}_2$.
Then, we apply lemma $2$ to decompose ${\cal B}_1$ into spaces
${\cal W}_1$ and ${\cal W}^{\perp}_1$ so that  $J_1$ looks like
\be
\label{j1decomp}
\begin{pmatrix}
... & O(\epsilon_4) \\
O(\epsilon_4) & ...
\end{pmatrix}
\ee
and similarly for $J_2$.
Inserting Eq.~(\ref{j1decomp}) into Eq.~(\ref{hj1j2}), $H$ looks like (in the
new basis)
\be
\begin{pmatrix}
... & O(\epsilon_4) & O(\epsilon_5) & O(\epsilon_3\epsilon_5) \\
O(\epsilon_4) & ... & ... & O(\epsilon_3) \\
O(\epsilon_5) & ... & ... & O(\epsilon_4) \\
O(\epsilon_3\epsilon_5)&O(\epsilon_3)& O(\epsilon_4) & ...
\end{pmatrix}
\ee
which is close to the block diagonal matrix,
\be
\begin{pmatrix}
... & & \\
& ... & ... & \\
& ... & ... & \\
&&  & ...
\end{pmatrix}
\ee
which has $3=n_{cut}+1$ blocks.

\section{Proof of Lemma \ref{mainlemma}}

Let the space ${\cal V}_1$ be $d_1$ dimensional, with
orthonormal basis vectors $v_1,...,v_{d_1}$.
Let $S$ denote the $D_B$-by-$d_1$ matrix whose columns
are these basis vectors, so that $S$ is an isometry.

Define a function $\ffn(\omega_0,r,w,\omega)$ as follows.
Let $\ffn(0,0,1,\omega)=1$ for $\omega=0$.  Let
$\ffn(0,0,1,\omega)=0$ for $|\omega|\geq 1$.  
Let $\ffn(0,0,1,\omega)=\ffn(0,0,1,-\omega)$.
For $0 \leq \omega \leq 1$,
choose $\ffn(0,0,1,\omega)$ to be infinitely differentiable
so that
the Fourier transform of $\ffn(0,0,1,\omega)$,
which we write $\tilde\ffn(0,0,1,t)$, is bounded by a function which
decays faster than any polynomial.
Finally, we impose $\ffn(0,0,1,\omega)+\ffn(0,0,1,1-\omega)=1$ for
$0\leq \omega \leq 1$.

For general $\omega_0,r,w,$ define
the function $\ffn(\omega_0,r,w,\omega)$ by
$\ffn(\omega_0,r,w,\omega)=1$ for $|\omega-\omega_0|\leq r$, and
$\ffn(\omega_0,r,w,\omega)=\ffn(0,0,1,(|\omega-\omega_0|-r)/w)$
for $|\omega-\omega_0|\geq r$.
Then $\ffn(\omega_0,r,w,\omega)=0$ for $|\omega-\omega_0|\geq r+w$.
For $r\geq 0$ and $w>0$, the function
$\ffn(\omega_0,r,w,\omega)$ is infinitely differentiable with
respect to $\omega$ everywhere.
The functions $\ffn(0,1,1,\omega)$ and $\ffn(0,0,1,\omega)$ are sketched
in Fig.~1a,b; the variable $r$ denotes the width of the flat part at the
center of the function, while $w$ denotes the width of the changing part of
the function.
Since $\ffn(0,0,1,\omega)$ is infinitely
differentiable, there is a function $T(x)$ which decays faster
than any polynomial such that:
\begin{eqnarray}
\int_{|t|\geq t_0} {\rm d}t |\tilde\ffn(\omega_0,w,w,t)| \leq T(wt_0),
\\ \nonumber
\int_{|t|\geq t_0} {\rm d}t |\tilde\ffn(\omega_0,0,w,t)| \leq T(wt_0).
\end{eqnarray}

The operator norm of $J$ is bounded by $1$.
The idea of the proof is to divide the
interval of eigenvalues of $J$, which is
$[-1,1]$, into various small overlapping
windows.  Then, for each interval centered on a frequency
$\omega$,
we will construct vectors given by approximately projecting vectors in
${\cal V}_1$ onto the space spanned by
eigenvectors of $J$ with eigenvalues lying in
that interval; we call the spaces of these vectors ${\cal X}_i$,
where $i$ labels the particular window.
Then, each of these projected vectors $x$
will have the property that $Jx$ is close to $\omega x$.
This will be the key step in ensuring property
{\bf (2)} in the claims of the lemma.
The idea of {\it approximate} projection is
important here.  In fact, we will use the smooth
filter functions $\ffn(\omega_0,r,w,\omega)$ above.
The smoothness will be essential to ensure
that the vectors $x$
have most of their amplitude in the first blocks
rather than the last blocks.  Since the vectors in the spaces ${\cal X}_i$
are approximate projections of vectors in ${\cal V}_1$ into different
windows,
we will be able to approximate
any vector $v_1 \in {\cal V}_1$ by
a vector in the space spanned by the ${\cal X}_i$ simply by
adding up the projections of $v_1$ in each different window.
Because the windows overlap, the vectors
may not be orthogonal to each other; the overlap
between vectors is something we will need to bound
(see Eq.~(\ref{eisomi}) below).  To control the overlap,
we choose ${\cal W}$ to be a subpace of the space spanned
by the ${\cal X}_i$ as explained below; this will then require us
to be careful to ensure that we are still able to approximate
vectors in ${\cal V}_1$ by vectors in ${\cal W}$.

Let $n_{win}$ be some even integer chosen later.
We will choose 
\be
\label{nwindef}
n_{win}=L/F(L),
\ee
where the function $F(L)$ is a function that grows slower than any power
of $L$ and is defined further below.  The choice of function $F(L)$ will
depend only on the function $T(x)$ defined above.

\begin{figure}
\label{ffn}
\centerline{
\includegraphics[scale=0.6,angle=0]{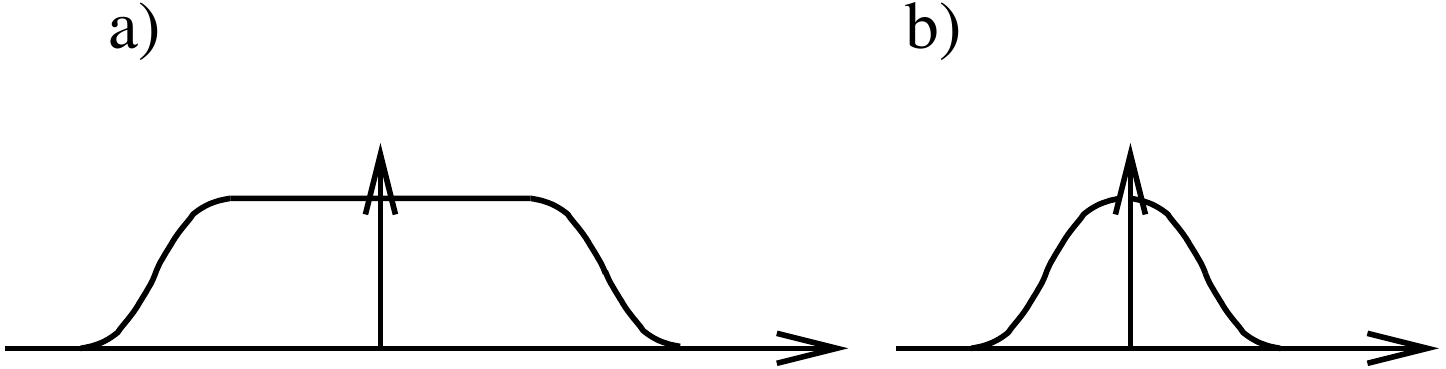}}
\caption{Sketch of (a) $\ffn(0,1,1,\omega)=\ffn(-1,0,1,\omega)+\ffn(0,0,\omega)+\ffn(1,0,1,\omega)$
and (b) $\ffn(0,0,1,\omega)$.}
\vspace{5mm}
\end{figure}

For each $i=0,...,n_{win}-1$,
define 
\be
\omega(i)=-1+2i/(n_{win}-1).
\ee
Define
\be
\fs=2/(n_{win}-1),
\ee
so $\omega(i)=-1+i\fs$.

When $\omega(i)$ and $\fs$ are chosen as above, we have
$\sum_{i=0}^{n_{win}-1} \ffn(\omega(i),0,\fs,\omega)=1$ for $-1\leq \omega \leq 1$.
See Fig.~2(a) to see a sketch of three functions
$\ffn(\omega(i-1),0,\fs,\omega)$,
$\ffn(\omega(i),0,\fs,\omega)$, and
$\ffn(\omega(i+1),0,\fs,\omega)$; as $\ffn(\omega(i),0,\fs,\omega)$
decreases for $\omega(i)\leq \omega \leq \omega(i+1)$,
the function $\ffn(\omega(i+1),0,\fs,\omega)$
is increasing to keep the sum constant.

\subsection{Construction of Spaces ${\cal X}_i$}

To construct ${\cal X}_i$, 
we
define the matrix $\tau_i$ by
\begin{eqnarray}
\label{und}
\tau_i &=&
\ffn(\omega(i),0,\fs,J)S.
\end{eqnarray}
Define
\be
\lambda_{min}=1/(n_{win}L^2).
\ee

\begin{figure}
\label{overlap}
\centerline{
\includegraphics[scale=0.6,angle=0]{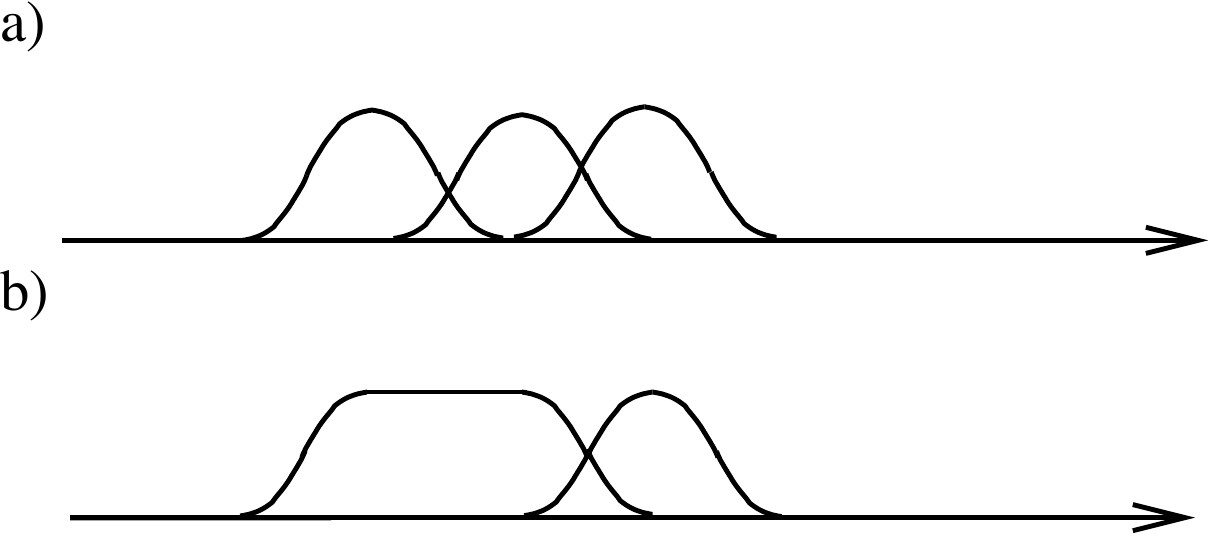}}
\caption{a) Sketch of overlapping windows. b) Re-arrangement of windows as
discussed in section on tridiagonal matrices.}
\vspace{5mm}
\end{figure}

Compute the eigenvectors of the matrix $\tau_i^{\dagger} \tau_i$.  For each
eigenvector $x_a$ with eigenvalue greater than or equal to
$\lambda_{min}$ compute $y_a=\tau_i x_a$.  Let
${\cal X}_i$ be the space spanned by all such vectors $y_a$.
Let $Z_i$ project onto the eigenvectors $x_a$ with eigenvalue less
than $\lambda_{min}$; the projector $Z_i$ will be used later in
computing the error estimates.

{\bf Remark:} To understand this construction, in Fig.~2a we
sketch the functions
$\ffn(\omega(i-1),0,\fs,\omega)$,
$\ffn(\omega(i),0,\fs,\omega)$, and
$\ffn(\omega(i+1),0,\fs,\omega)$, which
form partially
overlapping windows.
Note that the vectors
$\ffn(\omega(i),0,\fs,J)Sx_1$ and
$\ffn(\omega(i\pm 1),0,\fs,J)Sx_2$, for arbitrary $x_1,x_2$, need not be orthogonal.

\subsection{Properties of ${\cal X}_i$}

This subsection establishes certain properties of the ${\cal X}_i$.  It is primarily
intended to motivate the construction thus far.  We will show that the ${\cal X}_i$
have three properties which are closely related to the three properties we desire
to show in lemma (\ref{mainlemma}).

First, for any normalized vector $v\in {\cal V}_1$, the projection of $v$ onto
the orthogonal complement of the space spanned by the ${\cal X}_i$ is bounded
by $\sqrt{2n_{win}\lambda_{min}}=\sqrt{2}/L$.  To show this,
for any $v \in {\cal V}_1$, with $|v|=1$, we write
$v=Sx$ with $|x|=1$, and then
\begin{eqnarray}
\label{c1pre}
|v-\sum_{i=0}^{n_{win}-1}
\tau_i (1-Z_i) x |^2 &=&
|\sum_{i=0}^{n_{win}-1} \tau_i Z_i x|^2
\\ \nonumber
& \leq & 2 
\sum_{i=0}^{n_{win}-1} |\tau_i Z_i x|^2
\\ \nonumber
&\leq &
2 n_{win} \lambda_{min}
\\ \nonumber
& \leq & 2/L^2.
\end{eqnarray}
The factor of $2$ in the first inequality follows because
$(\tau_i Z_i x,\tau_j Z_j x)=0$ for $|i-j|>1$, but may be non-vanishing
for $i=j\pm 1$.  Similar factors of $2$ occur in
several other places.

Second, each space ${\cal X}_i$ is an approximate eigenspace of $J$.  That is,
for any $v_i \in {\cal X}_i$, we have
\be
\label{c2pre}
|(J-\omega(i))v_i|\leq \fs |v_i|.
\ee

Third, for any vector $y \in {\cal X}_i$, the norm of the
projection of $y$ onto ${\cal V}_L$ is bounded by $|y|$ times
a function growing slower than any power of $L$.  It is here that
we will pick the function $F(L)$ and
use the Lieb-Robinson bounds.  Let $\tilde\ffn(\omega_0,r,w,t)$
denote the Fourier transform of
$\ffn(\omega_0,r,w,\omega)$ with respect to the last variable $\omega$.
Then, for any $x$ with $(1-Z_i)x=x$, we find that
$y=\tau_i x=\ffn(\omega(i),0,\fs,J)Sx$ is equal to
\be
y=\int {\rm d}t \tilde\ffn(\omega(i),0,\fs,t) \exp(iJt) Sx.
\ee
We use the Lieb-Robinson bounds for matrix $J$, by defining a position matrix
which is equal to $i$ in the $i$-th block.
Using the Lieb-Robinson bounds, for time $t\leq L/\vlr$, with $\vlr=e^2$, we find that
the norm of the projection of
$\exp(iJt)Sx$ onto the space ${\cal V}_L$
is bounded by
$\exp(-L)$.  At the same time, the integral
$\int_{|t|\geq L/\vlr} {\rm d}t \tilde\ffn(\omega(i),0,\fs,t)$ is
bounded by $T(2 L/\vlr n_{win})=T(2 F(L)/e^2)$.
Since
$T(x)$ decays faster than any negative power of $x$,
we can choose an $F(x)$ which grows slower than
any power of $x$ such that $T(2 F(L)/e^2)$ still decays faster
than any negative power of $L$.
Thus, 
since $|y| \geq \lambda_{min} |x|$ 
by construction, 
for this choice of $F(x)$
the norm of the projection of any vector $y\in {\cal W}_i$ onto ${\cal V}_L$ is bounded by
$|y|$ 
times a function decaying faster than any negative power of $L$.

The reason for picking $\lambda_{min}>0$ is to help establish the third
property above.  Let us give an example of a situation where we would
encounter problems if we have taken $\lambda_{min}=0$.  Consider
a matrix of the form
\be
\begin{pmatrix}
0 & 1/4 \\
1/4 & 0 &1/4 \\
& 1/4 & 0 & 1/4 \\
&& 1/4 & 0 & 1/4 \\
&&& ... & \\
&&&& 1/4 & 0 & 1/4 \\
&&&&& 1/4 & 1/2
\end{pmatrix}
\ee
Here, each block has size one.
If it weren't for the ``$1/2$" in the last line, this matrix would have operator norm
slightly less than 
$1/2$.  However, because of the $1/2$, this matrix has one eigenvalue
greater than $1/2$.  For this particular choice of matrix, this eigenvalue
is close to $5/8$.
The corresponding eigenvector is localized near the last block, and is exponentially
small in the first block.  If we project a vector in ${\cal V}_1$ into a narrow window centered
on $\omega(i)=5/8$, the result will project onto this eigenvector, and thus the resulting
state will have large amplitude on ${\cal V}_L$.  However, for such a window, we would
find that $\tau_i$ would be exponentially small, and so we would not include this
vector in ${\cal X}_i$.

The properties we have established for spaces ${\cal X}_i$ are closely
related to the properties in lemma (\ref{mainlemma}) that we
are trying to establish.  Unfortunately, the spaces ${\cal X}_i$ need
not be orthogonal, and in fact may be very far from orthogonal.
This can lead to problems like the following: suppose we have two vectors,
$v_1 \in {\cal X}_1$ and $v_2 \in {\cal X}_2$.  We know that the projection
of $v_1$ onto ${\cal V}_L$ is small compared to $|v_1|$, and we know the
same thing for $v_2$; however, we don't know that the projection of
$v_1+v_2$ onto ${\cal V}_L$ is small compared to $|v_1+v_2|$ because
we don't know how $|v_1+v_2|$ compares to $|v_1|$ and $|v_2|$.
We have two different ways of dealing with this: in the next subsection,
we present a construction for block tridiagonal matrices that involves
choosing a subspace of the space spanned by the ${\cal X}_i$.  In
a later section on tridiagonal matrices, we present a much simpler
construction that involves combining several windows into one; the
reader may prefer to read that section first.

\subsection{Construction of ${\cal W}$}
We now construct the space ${\cal W}$.
Let each space ${\cal X}_i$ have dimension $D_i$.  
In each space ${\cal X}_i$ we can find an orthonormal basis of vectors,
$v_{i,b}$, for $b=1,...,D_i$.
We define a block tridiagonal matrix $\rho$ of inner products
of vectors $v_{i,b}$ as follows: the $i$-th block (for $0\leq i <n_{win}$) has dimension $D_i$,
and on the diagonal the matrix is equal to the identity matrix.  Above the
diagonal, the block in the $i$-th row and $i+1$-st column is equal to
the matrix of inner products $(v_{i,b},v_{i+1,c})$ for $b=1,...,D_i$ and
$c=1,...,D_{i+1}$.  
Note that
for $|i-j|>1$, the 
spaces ${\cal X}_i$ and ${\cal X}_j$
are orthogonal, so that the matrix $\rho$ is block tridiagonal.
We define a new vector space ${\cal R}$ to be a space of dimension
$\sum_{i=0}^{n_{win-1}} D_i$. 
The matrix $\rho$ is Hermitian and positive semidefinite.  It is equal to
$\rho=A^{\dagger} A$, for some matrix $A$ which is also block tridiagonal.
The matrix
$A$ is a linear operator from ${\cal R}$ to ${\cal B}$; it is simply
a matrix whose columns, in a given block, are different basis vectors for the
space ${\cal X}_i$ corresponding to that block.

{\bf Remark:} The matrix $\rho$ is block tridiagonal.  To motivate what
follows, consider the following circular reasoning: given that $\rho$ is
block tridiagonal, if we knew that theorem (\ref{btridthm}) were true,
we could find a basis in which $\rho$ was approximately diagonal and in
which a position operator, a block diagonal matrix
equal to $i/n_{win}$ in the $i$-th block,
was also approximately diagonal.  Then we choose ${\cal W}$ to
be the space spanned by vectors of the form $Aw_i$, where $w_i$
are basis vectors in this basis for which the
diagonal entry of $\rho$ are not too close to zero (how close is something we
would pick later).  Then, we would know that the vectors $Aw_i$ and
$Aw_j$ are not degenerate for $i \neq j$, and the operator $A$ would
be an approximate isometry from the space spanned by the $w_i$ to
${\cal W}$.  Also, we would find that $Aw_i$ was an approximate eigenvector
of $J$.  We would know that any vector $v \in {\cal V}_1$ had small projection
orthogonal to ${\cal W}$, since $v$ could be written as $Sx$ and, while
$x$ may have some projection onto vectors $w_i$ for which the diagonal
entry of $\rho$ is very close to zero, the error in  $v$ we make by
dropping those vectors from $x$ is small.
This would give the space ${\cal W}$ the properties we are trying
to construct.  Unfortunately, of course, we are trying to prove
theorem (\ref{btridthm}), so this line of reasoning does not help.
However, we do not need such a strong result in the present construction
as will be seen below.
Importantly, if there is a vector $w$ such that
$(w,\rho w)$ is small, it leads to only a small error in our ability to
approximate vector in $v \in {\cal V}_1$ if the vector $w$ is orthogonal to
the space ${\cal W}$.  We will also make use of a related fact:
if there is a vector $w=w_1+w_2$ such that $(w_1+w_2,\rho (w_1+w_2))$ is
small, then this means that $Aw_1$ is close to $-Aw_2$.  Suppose
$Aw_1\in {\cal X}_1$ and $Aw_2 \in {\cal X}_2$.  Then, we can
take ${\cal W}$ to be the space spanned by ${\cal X}_2,{\cal X}_3,...$ and
spanned by
the subspace of ${\cal X}_1$ orthogonal to $Aw_1$, and this
leads to only a small error in our ability to approximate vectors $v\in {\cal V}_1$
by vectors in ${\cal W}$.  
This is the basic idea behind the construction that follows.

We define spaces
${\cal Y}_i$, for $i=0,...,n_{sb}-1$, as follows, where $n_{sb}$ is the smallest
even integer greater than or equal to
$n_{win}/l_b$ with the ``block length" $l_b$ being an integer
equal to
\be
l_b =\lfloor n_{win}^{1/3} \rfloor.
\ee
Here, ``sb" stands for ``super-block" as we combine several blocks into one superblock.
We pick ${\cal Y}_i$ to be the subspace of ${\cal R}$
spanned by the vectors in blocks from the $(i-1)l_b$-th block to the $(i+1)l_b-1$-th block.
That is, it is the subspace spanned by vectors in blocks
$(i-1)l_b,(i-1)l_b+1,(i-1)l_b+2,...,(i+1)l_b-1$.
Therefore, ${\cal Y}_i$ is orthogonal to ${\cal Y}_j$ for $|i-j|>1$.
The space spanned by the ${\cal X}_i$ for $i=0,...,n_{win}-1$ is the same as the space spanned by
the $A{\cal Y}_i$ for $i=0,...,n_{sb}-1$; we will choose
the space ${\cal W}$ to be a subspace of this space.
Let $P_i$ project onto the subspace of
${\cal R}$
spanned by the blocks from the $il_b$-th block to the $(i+1)l_b-1$-th block.
For notational convenience later (and to avoid various off-by-one errors), we define
$P_{-1}=0$, and we define ${\cal X}_i$ for $i<0$ to be the empty set.

In Fig.~3 we sketch the blocks used to define the spaces ${\cal Y}_i$ for
the case $n_{sb}=6$.
The horizontal position in the figure indicates increasing block number, as
marked in the top row.  Space ${\cal Y}_i$ overlaps with space ${\cal Y}_{i\pm 1}$, as seen.  We have also sketched the range of the operators $P_i$.
\begin{figure}
\centerline{
\includegraphics[scale=0.6, angle=0]{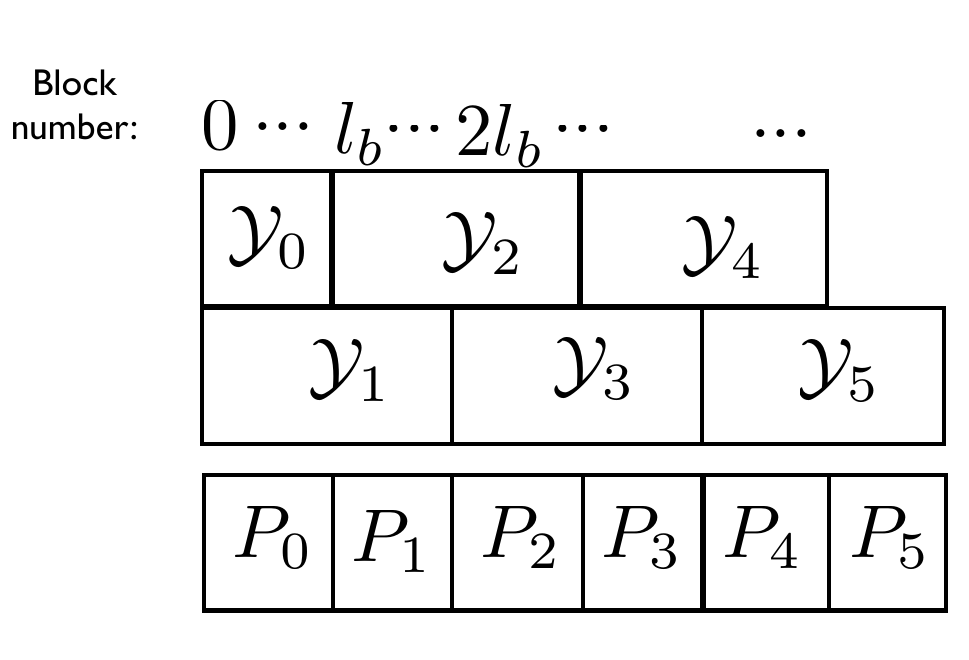}}
\caption{Sketch of which blocks are in which subspaces ${\cal Y}_i$ for
$n_{sb}=6$, as well
as which blocks are in the range of the $P_i$.}
\vspace{5mm}
\end{figure}

We will need to make use of the following lemma.  Currently our proof of this lemma relies on Lin's theorem, and thus
the proof is not independent of Lin's theorem.  However, we will only need the lemma below for  {\it some} $\epsilon<1$ (any such
$\epsilon$ suffices---note that for $\epsilon=1$ the result is trivial) and we hope that an independent proof can be given of the lemma for some $\epsilon<1$.  Thus, currently
our proof ``bootstraps" Lin's result, showing that once Lin's result holds for
some nontrivial $\epsilon$, then a polynomial dependence between $\epsilon$ and $\delta$ follows; indeed, all we need is the following
lemma which is a corollary of Lin's theorem so that this lemma is equivalent to Lin's theorem.

\begin{lemma}
\label{cutlemma}
For every $\epsilon>0$ there exists a $\delta>0$ such that the following holds for any matrices $A$ and $B$ with $\Vert A \Vert,\Vert B \Vert
\leq 1$ and $\Vert [A,B] \Vert \leq \delta$.  Let $P^-$ project onto the eigenspace of $A$ with eigenvalues $\leq -1/2$, let
$P^+$ project onto the eigenspace of $A$ with eigenvalues $\geq +1/2$.  Then, there exists a projector $P$ such that
the range of $P^-$ is a subspace of the range of $P$,
the range of $P^+$ is a subspace of the range of $1-P$, and
$\Vert [P,B] \Vert \leq \epsilon$.
\begin{proof}
By Lin's theorem, for every $\theta>0$, there exists a $\delta>0$ such that for every such $A$ and $B$ there exist operators
$A'$ and $B'$ with 
$\Vert A'-A \Vert \leq \theta$,
$\Vert B'-B \Vert \leq \theta$, and $[A',B']=0$.

Let $Q$ project onto the 
the eigenspace of $A'$ with eigenvalues in the interval $[-1/4,+1/4]$. 
Apply Jordan's lemma to the projector $Q$ and the projector $P^-+P^+$ to construct an orthonormal basis of vectors $q_i$ for the
range of $Q$ such that $(q_i,(P^-+P^+)q_j)=0$ for $i\neq j$.
Since $\Vert A'-A \Vert \leq \theta$, for every
such vector $q_i$ we have that
\be
(q_i,(P^-+P^+)q_i) \leq {\rm const.} \times \theta.
\ee

Define $Q'$ to project onto the space spanned by $(1-(P^-+P^+))q_i$ for all $i$.
Note that
\be
\label{qpqclose}
\Vert Q'-Q \Vert \leq {\rm const.} \times \theta.
\ee
We have defined three orthogonal projectors, $Q',P^-,P^+$.  Note that $\Vert [Q',B'] \Vert \leq {\rm const.} \times \theta$ since $[Q,B']=0$ and
Eq.~(\ref{qpqclose}) bounds $\Vert Q'-Q\Vert$.  Similarly, $\Vert [Q',A'] \Vert \leq {\rm const.} \times \theta$.
These bounds imply that $\Vert [Q,A] \Vert,\Vert[Q,B] \Vert$ are bounded bounded by ${\rm const.} \times \theta$.
Let $R=1-Q'$.
Then, $[R A R,R B R]$ is bounded by a constant times $\theta$ plus a constant times $\delta$.

We claim that, for sufficiently small $\theta$, the matrix $A$ projected into the range of $R$ has no eigenvectors with eigenvalue smaller than
$1/8$ in absolute value.  Since $\Vert A'-A \Vert$ is bounded, and
$\Vert Q'-Q \Vert$ is bounded by Eq.~(\ref{qpqclose}), this claim follows for sufficiently small $\theta$
since the matrix $A'$ projected into the
the range of $I-Q$ has no eigenvectors with eigenvalue smaller than $1/4$  in absolute value.

Now, we define the projector $P$ to be the projector $Q'$ plus the projector onto the negative eigenvalues of $R A R$.
(Equally well, we could define $P$ to just be the projector onto the negative eigenvalues of $RAR$.)

By construction, since $Q'$ is orthogonal to $P^-$ and $P^+$, the range of $P^-+P^+$ is a subspace of the range of $R$.
Thus, every eigenvector of $A$ with eigenvalue $\leq -1/2$ is an eigenvector of $RAR$ with eigenvalue $\leq -1/2$, and so
the range of $P^-$ is a subspace of the range of $P$.  Similarly,
the range of $P^+$ is a subspace of the range of $1-P$.  

Now, we have to bound $\Vert [P,B] \Vert$ which requires choosing $\theta$ sufficiently small and hence requires choosing $\delta$ sufficiently small.
By choosing $\theta$ sufficiently small, we can take $\Vert B'-B \Vert \leq \epsilon/4$, so it suffices to bound
$\Vert [P,B'] \Vert$ by $\epsilon/2$ in order to bound $\Vert [P,B] \Vert$ by $\epsilon$.  However, since $Q'$ almost commutes with $B'$ (that is, the commutator is bounded in norm by a constant times $\theta$ plus a constant times $\delta$), we only have to bound
the commutator $[P,B]$ in the space which is the range of $R$.  That is, it suffices to bound $\Vert [P,RBR] \Vert$.  Let $v$ be a vector which is in the range of $R$ and $P$ and let
$w$ be a vector which is in the range of $R$ and $1-P$.  We have to bound $(v,RBR w)$ for all such $v,w$.  However,
$RAR$ and $RBR$ almost commute (again, the commmutator is bounded in norm by a constant times $\theta$ plus a constant times $\delta$) and $v$ is in the eigenspace of $RAR$ with eigenvalues at most $-1/8$ and $w$ is in the eigenspace
of
$RAR$ with eigenvalues at least $1/8$ so we have bounded $(v,Bw)$ by a constant times $\theta$ plus a constant times $\delta$.
Thus, choosing $\theta$ sufficiently small (a constant amount smaller than $\epsilon$), and picking $\delta$ as given by  Lin's theorem
the claim follows.
\end{proof}
\end{lemma}

We claim that
\begin{lemma}
\label{nlemma}
There exist spaces ${\cal N}_i$, for $i=0,...,n_{sb}-1$ with the properties that:
\begin{itemize}
\item[{\bf 1}:] ${\cal N}_i$ is a subspace of ${\cal Y}_i$.  In fact, ${\cal N}_i$ is a subspace of the space spanned by
blocks with $(i-3/4)l_b\leq j <(i+3/4) l_b$.

\item[{\bf 2}:] For any vector $v \in {\cal N}^0_i$, the quantity $(v,\rho v)$ is bounded by
$|v|^2/l_b^2$ times a function $F_0(l_b)$, which is
growing slower than any power of $l_b$.

\item[{\bf 3}:] 
Let $N_i$ project onto ${\cal N}_i$.
For any vector $v$ which is in the space spanned by
eigenvectors of $\rho$ with eigenvalue
less than $1/l_b^2$, the sum $\sum_i |N_i v|^2$ is greater
than or equal to $(1-F_1(l_b)) |v|^2$, where
$F_1(l_b)$ is
a function decaying faster than any negative power
of $l_b$.

\item[{\bf 4}]: For any $i$, let $v$ be any vector in the space spanned by blocks with $(i-3/4) l_b \leq j < (i-1/4) l_b$.  Consider
the vector $N_i v$.  The projection of this vector  onto blocks $(i+1/4) l_b \leq j < (i+3/4) l_b$ is bounded in norm squared by
$|v|^2 (1/4-\chi)$, for some $\chi>0$.
\end{itemize}
\begin{proof}
Consider the matrix $\rho$
projected into blocks with $(i-3/4)l_b\leq j <(i+3/4) l_b$.  Call this matrix $\rho_i$.  The space ${\cal N}_i$ will be chosen
to contain the eigenspace of $\rho_i$ with eigenvalues less than or equal to $\lambda_0$ where $\lambda_0$ equals $1/l_b^2$ times a function growing slower than
any power of $l_b$.  It will also be chosen to be orthogonal to the eigenspace of $\rho_i$ with eigenvalues greater than or equal to
$\lambda_1$ where
$\lambda_1$ equals $1/l_b^2$ times a function $F_0(l_b)$, which is growing slower than any power of $l_b$.
As a result, {\bf 1,2} above are automatically satisfied for any such space ${\cal N}_i$.

We will now show that {\bf 3} also follows for any such ${\cal N}_i$.  Then, we will show how to pick ${\cal N}_i$ such that
{\bf 4} follows (here we use lemma (\ref{cutlemma}).

To show {\bf 3}, 
define a matrix $M$ by
\be
M=\begin{pmatrix}
0 & A \\
A^{\dagger} & 0
\end{pmatrix}.
\ee
Define a matrix $O$ by
\be
O=\int {\rm d}t \, \exp(i M t) \tilde \ffn(0,G(l_b)/l_b,G(l_b)/l_b,t),
\ee
where $G(l_b)$ is a function growing slower than any power of $l_b$
to be chosen later.
Note that since $\ffn$ is even in $t$, $O$ is non-vanishing only in the
upper left and lower right corners.
We define $O_0$ to be the top left corner of $O$.
For each $i$, define $A_i$ to project onto the blocks with
$(i-1/2)l_b\leq j <(i+1/2) l_b$.
Define $A'_i$ to project onto the blocks with $(i-3/4)l_b\leq j <(i+3/4) l_b$.

Let
$v$ be a normalized vector in the eigenspace of $\rho$ with eigenvalue
less than $1/l_b^2$. We can write $v=\sum_i a_i v_i$, with
$v_i$ a normalized vector in the space projected onto by $A_i$ and
the $a_i$ are complex amplitudes such that $\sum_i |a_i|^2=1$.
Then, since $O_0v=v$, we have $\sum_i |(O_0 v_i,v)|^2=|v|^2=1$.
Note that the vector $O_0 v_i$ is in the eigenspace of $\rho$ with eigenvalue less than or equal to $2G(l_b)/l_b$.
Further, by using the Lieb-Robinson bounds, the
quantity
$|A'_i O_0 v_i - O_0 v_i|$ can be made to go to zero faster than
any power of $l_b$.  In fact, by using the Lieb-Robinson bounds one
can ensure that if we define $M_i$ by
\be
M_i=\begin{pmatrix}
0 & A'_i A A'_i \\
A'_i A^{\dagger} A_i & 0
\end{pmatrix},
\ee
and define $O_0^i$
to be the top-left corner of the matrix $\int {\rm d}t \, \exp(i M_i t) \tilde \ffn(0,G(l_b)/l_b,G(l_b)/l_b,t)$,
then
$|O_0^i v_i - O_0 v_i|$ can be made to go to zero faster than any power of $l_b$ (note that
the matrix $O_0^i$ is non-vanishing only in blocks with
$(i-3/4)l_b\leq j <(i+3/4) l_b$; the smallness of this difference here is a statement that due to the Lieb-Robinson bounds, changing
the effects near the boundary of the block only weakly affects the dynamics).
Note, however, that the vector $O_0^i v_i$ is in the eigenspace of $\rho_i$ with eigenvalue $\leq 2 G_0(l_b)/l_b$.
Hence,
by choosing $\lambda_0= 2 G(l_b)/l_b$,
one can ensure that
up to error which goes to zero faster than any power of $l_b$,
the vector $O_0 v_i$ is in the eigenspace of $\rho_i$ with eigenvalues less than or equal to $\lambda_0$, and hence
the vector $O_0 v_i$ is in ${\cal N}_i$, up to error going to zero faster than any power of $l_b$.
Hence,
$|N_i v|^2$ is greater than or equal to $|(O_0 v_i,v)|^2$ minus
a quantity going to zero faster than any power of $l_b$ (to see this, note that $|N_i v|^2\geq |(x,v)|^2$ for any vector
$x$ in ${\cal N}_i$ and then we use the fact that $O_0 v_i$ is in ${\cal N}_i$ up to error going to zero faster than any power
of $l_b$).
Thus, since $\sum_i |(O_0 v_i,v)|^2=\sum_i |(v_i,O_0 v)|^2=|v|^2=1$,
this verifies {\bf 3}.

We finally show {\bf 4}.  We can show this by applying lemma (\ref{cutlemma}), with $B$ being a block identity matrix which is
equal to $-1$ on blocks
$(i-3/4) l_b \leq j < (i-1/4) l_b$, equal to $+1$ on blocks
$(i+1/4) l_b \leq j < (i+3/4) l_b$, and linearly interpolating in between.  We define $A$ to be a function of $\rho_i$: we define
a smooth function $f(x)$ such that $f=-1$ for $x\leq \lambda_0$, $f=+1$ for $x\geq \lambda_1$, and $f$ smoothly interpolates.
Then, we can bound the commutator $[A,B]$ by a quantity going to zero faster than any power of $l_b$.
Hence, appylying lemma (\ref{cutlemma}), we can choose a projector $P$ with the property that $\Vert [P,B] \Vert$ is bounded by a quantity strictly less than $1$.  Hence, we obtain {\bf 5}.
\end{proof}
\end{lemma}

{\bf Remark:} The definition of $M$ and $O$ as block matrices in the
above lemma is simply a trick to make the claims ${\bf 2,3}$ in the
lemma depend on $l_b^{-2}$ rather than $l_b^{-1}$ as we would have found
without this trick of introducing block matrices.  In physics jargon,
near the edge of the band (eigenvalues close to zero for $\rho$ which is
a positive semi-definite matrix), we have dynamic critical exponent
$2$ rather than $1$.

We have chosen the overlapping of the blocks in a particular way.  A vector in ${\cal N}_i$ may have support on blocks $(i-3/4)l_b \leq j
< (i+3/4) l_b$.  Thus, on blocks $(i-3/4) l_b \leq j <(i-1/4) l_b$, it may overlap with a vector on ${\cal N}_{i-1}$ and on blocks
$(i+1/4) l_b \leq j <(i+3/4) l_b$, it may overlap with a vector on ${\cal N}_{i+1}$.

We now describe the construction.  The numeric constant $\eta$ below
is some sufficiently small, positive, real number; the choice of this
number will be discussed in lemma (\ref{decaylemma}).

We iteratively construct a sequence of spaces ${\cal N}'_i$ for odd $i$ which are subspaces
of ${\cal N}_i$ as follows.
For each $i=1,3,...$,
consider the space ${\cal N}_i$.
Let $Q_i$ be the projector
onto the span of ${\cal N}_{i+1},{\cal N}_{i-1}$.
Apply Jordan's lemma\cite{jl} to $N_i$
and to $Q_i$
to construct a
complete orthonormal basis $n_{i,b}$ for ${\cal N}_i$ such that
$(n_{i,b},Q_i n_{i,c})=0$ for $b\neq c$.  Let ${\cal N}'_i$ be the
space spanned by vectors $n_{i,b}$ such that $|Q_i n_{i,b}|^2\leq 1/2+\eta$.
Define ${\cal U}^\perp$ to be the space spanned by
the ${\cal N}_i$ for even $i$ and by the ${\cal N}'_i$ for odd $i$.
Define ${\cal U}$ to be the subspace of ${\cal R}$ orthogonal to ${\cal U}^\perp$.
We now define ${\cal W}=A {\cal U}$.
Let $P$ be the projector onto ${\cal W}$.  We define $U$ to be
the projector onto ${\cal U}$.  Thus, for any vector
$w$ with $Pw=w$, we have $w=Av$, for some $v$ with $U v=v$.

We claim one important property of this space ${\cal U}$:
\begin{lemma}
\label{decaylemma}
Let $\eta$ be a sufficiently small positive number.
Let $x_i$ be any vector in ${\cal Y}_i$.
Consider the vector
$U x_i$.  Project this vector into space ${\cal Y}_j$.
The norm of the resulting vector is bounded by
\be
\label{decaybnd}
{\rm const}. \times \exp(-|i-j|/{\rm const}.) |x_i|,
\ee
for some positive numeric constants.
\begin{proof}
We consider instead the vector $(1-U)x_i$ and bound the norm of the projection of that vector.
Since the space ${\cal U}^\perp$ is the span of spaces ${\cal N}_j$ for $j$ even
and ${\cal N}'_j$ for $j$ odd,
the projection $(1-U)x_i$ can be computed by minimizing the quantity
\be
\label{qf}
|\sum_j a_j n_j - x_i|^2
\ee
over all $|n_j|$ in ${\cal N}_j$ for $j$ even and $n_j$ in ${\cal N}'_j$ for $j$ odd, with
$|n_j|=1$, and over all complex numbers $a_j$.  Then, $(1-U)x_i=\sum_j a_j n_j$.
For given $x_i$, let the minimum be obtained for some definite  choice of vectors $n_j$.
Then, we consider the minimum over $a_j$ of (\ref{qf}).
We can write Eq.~(\ref{qf}) in a matrix form, by
introducing a tridiagonal matrix $M$, with diagonal entries equal to unity and entries $M_{i,i+1}=(n_i,n_{i+1})$.
Then, define a vector $\vec x$ which has its $j$-th entry equal to
$(n_j,x_i)$.
Define a vector $\vec a$, with $j$-th entry equal to $a_j$.
Then, the minimum over $a_j$ is given by
\be
\vec a = M^{-1} \vec x.
\ee

We will prove an exponential decay on entries of $M^{-1}$.  
That is,
we will define $G=M^{-1}$ and prove that the
matrix element $G_{ij}$ decays exponentially in $|i-j|$.  Then,
since the only non-zero entries of $\vec x$ are the $i-1$,$i$, and $i+1$
entries, this will prove the desired result (\ref{decaybnd}).

For odd $i$, the vector $n_i$ is
a sum of two vectors, one supported on blocks
$(i-1)l_b\leq j< il_b$ and the other supported on blocks
$i l_b\leq j <(i+1) l_b$.  Let $c_i$ denote the norm of the first vector, and $d_i$ denote the norm
of the second vector,  so that $c_i^2+d_i^2=1$.
For each odd $i$, let $m_i$ denote the vector
$n_i$ projected orthogonal to space ${\cal N}_{i-1}$ and ${\cal N}_{i+1}$.
Note now that $m_i$ is not orthogonal to $m_{i\pm 2}$.  We do not normalize the vector $m_i$.

Suppose $i$ is odd.  To determine the decay of $G_{ij}$, we can project $i$ into the space
spanned by the $m_i$.  This projection is a vector $\sum_i a'_i m_i$, and the $a'_i$ can be determined
by the inverse $G'$ of a Hermitian matrix $M'$ which has matrix elements
$M'_{i,i}=|m_i|^2$ and $M'_{i+2,i}=M'_{i,i+2}=(m_i,m_{i+2})$ for odd $i$ and vanishes for even $i$.
To prove the exponential decay of $G'$, it suffices to show a lower bound $x$ on the smallest eigenvalue of $M'$.

By construction, $|M_{i,i+1}|^2+|M_{i,i-1}|^2 \leq 1/2+\eta$.  So,
$M'_{i,i}\geq 1/2-\eta$.
Note that $|M'_{i,i+2}|^2 \leq  |d_i|^2 |c_{i+2}|^2 (1/4-\chi)$.  We will now show a lower bound on the smallest eigenvalue of $M'$;
this will be based on the lower bound on the diagonal elements and  the upper bound on the off-diagonal elements.

Consider the matrix $M''=(1+x) (1/2-\eta)^{-1} M'$, for some small real number $x$.  This matrix has diagonal entries greater than or equal to $1+x$.
It has matrix element $|M''_{i,i+2}|^2 \leq |d_i|^2 |c_{i+2}|^2 (1+x)(1/4-\chi) (1/2-\eta)^{-2}$.  For small enough $\eta$ and $x$,
$(1/4-\chi)(1/2-\eta)^{-2}$ is bounded by $1$, for some positive constant (the exact value of the constant
depends on $\chi$).  
Let $O$ be the following Hermitian matrix: $O_{i,i}=1$ for odd $i$ and $O_{i+2,i}=O_{i,i+2}=d_i c_{i+2}$  All other entries vanish.
Note that $O$ is positive semi-definite by construction.  The matrix $M''$ has its off-diagonal elements bounded by those of $O$ and
its diagonal elements are at least $1+x$ so that the smallest eigenvalue of $M''$ is at least $x$.

Given a lower bound on the smallest eigenvalue of $M''$, the exponential decay follows.
\end{proof}
\end{lemma}

\subsection{Properties of ${\cal W}$}

In this section we establish certain properties for the space
${\cal W}$.  The main results are 
Eq.~(\ref{eisomi}),
controlling the overlap between vectors in this space,
and Eq.~(\ref{inbec}), showing that
for any vector $v$ in the space spanned by ${\cal X}_i$ with
$v=Ax$, the vector $Pv \in {\cal W}$ is close to $v$, where the
maximum distance $|Pv-v|$ between the vectors depends on $|x|$.

{\it First Property---}
By construction, for
any vector 
$r \in {\cal U}$, with
$|r|=1$,
we have
\be
\label{bnd}
(r,\rho r)\geq
{\rm const.} \times (1/l_b^2),
\ee
for sufficiently large $l_b$.

To show Eq.~(\ref{bnd}), we bound the inner product between
$r$ and $w$ for $w$ in the span of eigenvectors of $\rho$
with eigenvalue less than or equal to $1/l_b^2$.  Any such
$w$ can be written as a linear combination of vectors
$w^{even}$ in the span of ${\cal N}_i$ for even $i$ and
$w^{odd}$ in the span of ${\cal N}_i$ for odd $i$.
Note that $w^{even}$ is in $Q^\perp$.  By {\bf 4} in lemma (\ref{nlemma}),
the inner product $(w^{even},w^{odd})$ is greater than or equal
to minus a quantity going to zero faster than any power of $l_b$.
Thus, $|(1-U)w|^2$ is greater than or equal to $|w^{even}|^2$ minus
a quantity going to zero faster than any power of $l_b$.

Further, we write $w^{odd}=w_1+w_3$ where $w_1=\sum_{i=1,5,9,...} n_i$ and
$w_3=\sum_{3,7,11,...} n_i$, with $n_i \in {\cal N}_i$.
Note that by construction, each $n_i$ has projection at least $(1/2+\eta)|n_i|^2$
onto the span of spaces ${\cal N}_{i-1},{\cal N}_{i+1}$, so $w_1$ has
projection at least $(1/2+\eta) |w_1|^2$ onto the span of these spaces, and
$w_3$ has
projection at least $(1/2+\eta) |w_3|^2$ onto the span of these spaces.
Since $(w_1,w_3)=0$, we
have $|(1-U) (w_1+w_3)|^2 \geq |(1-U)w_1|^2+|(1-U)w_3|^2-2|(w_1,Uw_3)|
\geq |(1-U)w_1|^2-|(1-U)w_3|^2-2|Uw_1| |Uw_3|$.  Since $2|Uw_1||Uw_3|\leq
|Uw_1|^2+|Uw_3|^2$, 
$|(1-U) (w_1+w_3)|^2 \geq |(1-U)w_1|^2+|(1-U)w_3|^2-|Uw_1|^2-|Uw_3|^2\geq
2\eta (|w_1|^2+|w_3|^2)$ and so
$|(1-U) w|^2 \geq 2\eta |w|^2$
 minus a quantity going to
zero superpolynomially in $l_b$.
Therefore, having upper bounded $|U w|$, we have
upper bounded $(r,w)$, so Eq.~(\ref{bnd}) follows.

For any $v\in {\cal W}$, we can
find $x_i\in {\cal X}_i$ for $i=0,...,n_{win}-1$, such that $v=\sum_{i} x_i$.
Therefore, from Eq.~(\ref{bnd}), for any $v\in {\cal W}$, we can find $x_i$, $i=0,...,n_{win}-1$ with
$x_i \in {\cal X}_i$ and $v=\sum_i x_i$ with
\begin{eqnarray}
\label{eisomi}
|v|^2 \geq
{\rm const.} \times (1/l_{b})^2 \sum_{i=0}^{n_{win}-1} |x_i|^2.
\end{eqnarray}

{\it Second Property---}
We also claim that for any vector $v$ in the space spanned by the
${\cal X}_i$, such
that $v=Ax$
that
\be
\label{inbec}
|Pv-v| \leq {\rm const.} \times (\sqrt{F_0(l_b)}/l_{b}) |x|.
\ee
To show Eq.~(\ref{inbec}), any vector $x$ can be written
as a linear combination of a vector in ${\cal Q}$ and a vector
in $x^\perp \in {\cal Q}^\perp$.  Let $x=\sum_i x_i$, with $x_i\in {\cal Y}_i$
and $\sum_i |x_i|^2=|x|^2$.  The vector $x^\perp=(1-U) x$.
Let $x^{\perp}=\sum_j a_j n_j$, with $n_j$ in ${\cal N}_j$ for $j$ even
and $n_j$ in ${\cal N}'_j$ for $j$ odd as in lemma (\ref{decaylemma}).  
We bound
$|A x^\perp|^2$ by ${\rm const.} \times (F_0(l_b)/l_{b}^2) \sum_{|j-k|\leq 1}
|n_j| |n_k| \leq 
{\rm const.} \times (F_0(l_b)/l_{b}^2) \sum_{j} |n_j|^2 \leq
{\rm const.} \times (F_0(l_b)/l_{b}^2) \sum_{j} |x_j|^2$, where the last
inequality uses the exponential decay on matrix elements of $G$ from lemma (\ref{decaylemma}).

Any vector $v=\sum_{i=0}^{n_{win}-1} x_i$ with $x_i \in {\cal X}_i$
can be written as $v=Ax$ with $|x|^2 =\sum_{i=0}^{n_{win}-1} |x_i|^2$.
Therefore, Eq.~(\ref{inbec}) implies that for
any $v=\sum_{i=0}^{n_{win}-1} x_i$,
with $x_i\in {\cal X}_i$, we have
\be
\label{inbec2}
|Pv-v| \leq
{\rm const.} \times (\sqrt{F_0(l_b)}/l_{b}) 
\sqrt{\sum_{i=0}^{n_{win}-1} |x_i|^2}.
\ee

\subsection{Verification of Claims}
We now verify the claims regarding the subspace ${\cal W}$.

{\it Proof of First Claim---}
To prove ${\bf (1)}$, note that for any vector $v \in {\cal B}$ we have
\begin{eqnarray}
v&=&\sum_{i=0}^{n_{win}-1} 
\ffn(\omega(i),0,2n_{win},J)v.
\end{eqnarray}
For any $v \in {\cal V}_1$, with $|v|=1$, we can write
$v=Sx$ with $|x|=1$, and then, from Eq.~(\ref{c1pre})
\begin{eqnarray}
\label{c1}
|v-\sum_{i=0}^{n_{win}-1}
\tau_i (1-Z_i) x |^2
& \leq & 2/L^2.
\end{eqnarray}
The vector 
$\tau_i (1-Z_i) x$ is in ${\cal X}_i$.  So,
by Eq.~(\ref{inbec2}),
\begin{eqnarray}
\label{c2}
|(1-P) \sum_{i=0}^{n_{win}-1} \tau_i (1-Z_i) x|
& \leq  &
{\rm const.} \times (\sqrt{F_0(l_b)}/l_{b})
\sqrt{\sum_{i=0}^{n_{win}-1} | \tau_i (1-Z_i) x|^2}
\\ \nonumber
&\leq&
{\rm const.} \times (\sqrt{F_0(l_b)}/l_{b})
\sqrt{\sum_{i=0}^{n_{win}-1} | \tau_i x|^2}
\\ \nonumber
&\leq&
{\rm const.} \times (\sqrt{F_0(l_b)}/l_{b}).
\end{eqnarray}
Combining Eqs.~(\ref{c1},\ref{c2}) with
a triangle inequality verifies the first claim, given that
$F(L)$ is chosen to grow slower than any power of
$L$.

{\it Proof of Second Claim---}
To prove the second claim {\bf (2)},
consider any vector $v \in {\cal W}$. 
We have $v=\sum_i A x_i$, with $x_i \in {\cal X}_i$ and
$U \sum_i x_i=\sum_i x_i$,
so $v=\sum_i A U x_i$.  So,
$|(1-P) J v|=|\sum_i (1-P) J A  U x_i|$.  
We have $|\sum_i (1-P) J A  U x_i|^2=
\sum_{i,j} (JAU x_i,(1-P) J A U x_j)$.
Let $R_k$ project onto the $k$-th block of the space ${\cal R}$.
Note that $(1-P) A U x_i=0$, so $(1-P) \omega(il_b)A U x_i=0$, so
\begin{eqnarray}
|\sum_i (1-P) J A U x_i|^2  &=&
|(1-P) \sum_i\Bigl(J-\omega(il_b)\Bigr) A U x_i)|^2 \\ \nonumber
&\leq &
|\sum_i \Bigl(J-\omega(il_b)\Bigr) AU x_i|^2 \\ \nonumber
&\leq &
\sum_{i,j} |(\Bigl(J-\omega(il_b)\Bigr) AU x_i,\Bigl(J-\omega(jl_b)\Bigr) A U x_j)|.
\end{eqnarray}
Using the decay in lemma (\ref{decaylemma}), the inner product above
decays exponentially in $|i-j|$, so
we can sum over $i,j$ to find
\be
\label{p1}
(Jv,(1-P) J v) \leq {\rm const.} \times (l_b \fs)^2 \sum_i |x_i|^2.
\ee

By Eqs.~(\ref{p1},\ref{eisomi}), 
we have
\begin{eqnarray}
\label{f2}
|(1-P) J v|^2 
& \leq & {\rm const.} \times (l_b \fs)^2 \sum_i |x_i|^2
\\ \nonumber
& \leq & {\rm const.} \times 
(1/l_{b}^2)
\Bigl( l_b \fs )\Bigr)^2 |v|^2
\\ \nonumber
&=& {\rm const.} \times
\Bigl( l_b^2 \fs \Bigr)^2 |v|^2,
\end{eqnarray}
verifying the second claim.

{\it Proof of Third Claim---}
As we established before, using the Lieb-Robinson bound, for the given
choice of $F(x)$
the norm of the projection of any vector $y\in {\cal X}_i$ onto ${\cal V}_L$ is bounded by
$|y|$ times a function decaying faster than any negative power of $L$.
Let $P_{{\cal V}_L}$ project onto ${\cal V}_L$.
Using Eq.~(\ref{eisomi}), we find that 
the
projection of any vector $v \in {\cal W}$ onto
${\cal V}_L$ is bounded by (writing $v=\sum_{i=0}^{n_{win}-1} w_i$ with
$w_i\in {\cal X}_i$)
\begin{eqnarray}
|P_{{\cal V}_L} v|^2 & \leq &
n_{win} \sum_{i=0}^{n_{win}-1} 
|P_{{\cal V}_L} w_i|^2
\\ \nonumber
& \leq &
n_{win}
\, {\rm max}_i 
(|P_{{\cal V}_L} w_i|^2/|w_i|^2)
(1/l_{b})^2 |v|^2.
\end{eqnarray}
Since
$(|P_{{\cal V}_L} w_j|^2/|w_j|^2)$ is bounded by a
function decaying faster than any negative power of $L$, this
verifies the third claim.

This completes the proof of Lemma (\ref{mainlemma}).  After giving the error bounds
in the next section, we explain some of the motivation behind the
above construction, and comment on the easier case in which $J$ is
a tridiagonal matrix, rather than a block tridiagonal matrix.

\section{Error Bounds}
\label{erbnds}

We finally give the error bounds to obtains theorems (\ref{mainthm},\ref{btridthm}).
To obtain (\ref{btridthm}), we pick 
\be
n_{cut}=\Delta^{-1/4},
\ee
so
that $L=\lfloor (2/n_{cut})/\Delta)-1\rfloor$ is of order $2/\Delta^{3/4}$.
Then, from lemma 
(\ref{mainlemma}) and Eq.~(\ref{sumerr}), in the new basis the block-off-diagonal
terms in $H$ are bounded in operator norm by a constant times $\Delta^{1/4}$
times a function growing slower than any power of $1/\Delta$.
By Eq.~(\ref{bdiff}),
the difference between
$B$ and $B'$ is bounded in operator norm by a constant times $\Delta^{1/4}$.
Therefore, 
theorem (\ref{btridthm}) follows.
To obtain theorem (\ref{mainthm}), we pick
\be
\Delta=\delta^{4/5}
\ee
in lemma (\ref{h0con}).

We omit the detailed analysis, but it is possible to choose $E(x)$ to
be a polylog as follows.  We can pick $T(x)$ to decay like $\exp(-x^\eta)$,
for any $\eta<1$\cite{lim1,lim2}.
Then we can pick $F(L)$ to equal $\log(L)^{\theta}$, for
$\theta>1/\eta$, so that $T(F(L))\sim \exp(-(\log(L))^{\theta/\eta})$
decays faster than any power.

\section{Tridiagonal Matrices}
\label{motiv}

In this section, we
present tighter bounds for the case in which $H$ is a tridiagonal matrix,
rather than a block tridiagonal matrix.

{\bf Remark:}
The difficulty we face is that the ${\cal X}_i$ are
not orthogonal to each other.  If they were orthogonal, then many of the
estimates would be easier.  Consider the case in which $J$ is a block
diagonal matrix, so that ${\cal V}_1$ is one dimensional.  
Let $\rho(E)$ be a smoothed density of states
at energy $E$: $\rho(E)={\rm tr}(S^{\dagger} \ffn(E,1/L,1/L,J)^{\dagger}
\ffn(E,1/L,1/L,J) S)$.  Suppose
$\rho(E)$ is such that it has a peak in the crossing points of Fig.~2a (the points
where one function $\ffn$ is decreasing and the other is increasing and they cross).
Then, with the overlapping
windows as shown, we find that most of the smoothed density of states
lies in the overlap between the windows, rather than in the windows themselves.
The overlap between the vectors in different windows is large.
In the case of a tridiagonal matrix, we can combine two of the windows
as
shown in Fig.~2b to reduce the overlap of the normalized vectors; this
general idea will motivate the construction in this section.

We prove that
\begin{lemma}
\label{tridlemma}
Let $J$ be an $L$-by-$L$ Hermitian tridiagonal matrix, with $\Vert J \Vert \leq 1$
acting on a space ${\cal B}$.
Let  $v_j$ denote the vector with a $1$ in the $j$-th entry and zeroes elsewhere.
Then, there exists a space ${\cal W}$ which is a
subspace of ${\cal B}$ with the following properties:
\begin{itemize}
\item[{\bf (1)}:]
The projection of $v_1$
onto
the orthogonal complement of ${\cal W}$ has
norm bounded by $\epsilon_3$ where
$\epsilon_3$ is equal to a constant times $1/L$.

\item[{\bf (2)}:]
For any normalized vector
$w\in W$, the projection of
$J w$ onto
the orthogonal complement of ${\cal W}$ has
norm bounded by $\epsilon_4$, where
$\epsilon_4$ is equal to $1/L$ times
a function
growing slower than any power of $L$.

\item[{\bf (3)}:]
The projection of $v_L$ onto
${\cal W}$ has
norm bounded by $\epsilon_5$, where
$\epsilon_5$ is 
a function decaying faster
than any power of $L$.
\end{itemize}
\end{lemma}

This lemma implies theorem (\ref{tridtheorem}): we construct
$A',B'$ as before, following steps {\bf (3)} to construct
the new basis, but because
of the tighter bounds in lemma (\ref{tridlemma}) we can
choose $n_{cut}=\Delta^{-1/2}$ when constructing the new basis.
Now, in step {\bf (4)}, we find that $A',B'$ are diagonal
matrices, rather than just block diagonal matrices.

For each $i=0,1,...,n_{win}-1$,
define 
\be
\omega(i)=-1+i\fs,
\ee
as before.
Define
\begin{eqnarray}
\rho_{i}&=&{\rm tr}\Bigl(S^{\dagger} 
\ffn(\omega(i),0,\fs,J)^{\dagger}
\ffn(\omega(i),0,\fs,J) 
S\Bigr)
\\ \nonumber
&=& \Bigl|\ffn(\omega(i),0,\fs,J) v_1\Bigr|^2.
\end{eqnarray}
Set 
\be
\lambda_{min}=1/(n_{win}L^2),
\ee
as before
with
\be
n_{win}=L/F(L)
\ee
as before.
To prove Lemma (\ref{tridlemma}), we
use the following algorithm.  
There are $n_{win}$ windows, labeled $0,...,n_{win}-1$.
We label various windows as either ``unmarked" or ``marked"; windows which 
are marked get marked by an integer label.
\begin{itemize}
\item[{\bf 1:}] Set $i=0$.  Initialize a real variable $x$ to $0$.  
Initialize an integer counter $a$ to $1$.
Initialize all windows to unmarked.

\item[{\bf 2:}] Set $x$ to $0$.  {\bf If} $\rho_i< \lambda_{min}$,
\begin{itemize}
\item[{\bf then}]

\item[{\bf 2a:}] Increment $i$ by one.

\item[{\bf 2b:}] If $i\geq n_{win}$, terminate.  Otherwise, go to step {\bf 2}.

\item[{\bf endif}]
\end{itemize}

\item[{\bf 3:}] Mark window $i$ with label $a$.

\item[{\bf 4:}] Set $x$ to $x+\rho_i$.  {\bf If} $x<9 \rho_{i}$,
\begin{itemize}
\item[{\bf then}]

\item[{\bf 4a:}] Increment $i$ by one.

\item[{\bf 4b:}] If $i\geq n_{win}$, terminate.  Otherwise, go to step {\bf 3}.

\item[{\bf endif}]
\end{itemize}

\item[{\bf 5:}] Increment $a$ by one.  Increment $i$ by one.  If $i\geq n_{win}$, terminate.
Otherwise, goto step {\bf 2}.
\end{itemize}

After running this algorithm, there will
be a sequences of marked windows all marked with the same integer label $a$.
There may be
one or more unmarked windows separating the sequences of marked windows.
In step $2$, we scan along to find an $i$ with $\rho_i\geq \lambda_{min}$, and
then in step $4$ we mark a sequence of windows.
We claim that the length of a sequence of marked windows is at
most $1+\lceil \log_{10/9}(2/\lambda_{min})\rceil$.  This bound on the length of a sequence of marked windows holds because
at the start of a sequence
$x$ is at least $\lambda_{min}$, $x$ grows exponentially along
the sequence (otherwise in step {\bf 4} we find that
$\rho_{i+1}>(1/9) x$ for some $i$), and $x$ can be at most $2$ since $\sum_{i=0}^{n_{win}-1} \rho_i \leq 2$.

Let the total number of sequences be $n_{seq}$.  Note that $n_{seq} \leq n_{win}$.

For each sequence of windows marked with a given integer $a$, from
window $i$ to $j$, construct the vector $y_a$ given by
\begin{eqnarray}
y_a &=& \sum_{k=i}^j
\ffn(\omega(k),
0,\fs,J)v_1.
\\ \nonumber
&=& \ffn((\omega(i)+\omega(j))/2,
(\omega(j)-\omega(i))/2,\fs,J)v_1.
\end{eqnarray}
The inner product $(y_a,y_{a+1})$ is equal to
$(\ffn(\omega(j),0,\fs,J)v_1,y_{a+1})$.  By Cauchy-Schwarz,
this is bounded by 
$|(\ffn(\omega(j),0,\fs,J)v_1| |y_{a+1}|$.
To estimate 
$|(\ffn(\omega(j),0,\fs,J)v_1|$, we use
$|(\ffn(\omega(j),0,\fs,J)v_1|^2=\rho_j \leq
\sum_{k=i}^j \rho_k/9\leq
|y_a|^2/9=
\sum_{k=i}^j \sum_{k'=i}^j
(\ffn(\omega(k),0,\fs,J)v_1,
\ffn(\omega(k'),0,\fs,J)v_1)/9$,
where the first inequality is by construction
and the second inequality follows
from the fact that
$(\ffn(\omega(k),0,\fs,J)v_1,
\ffn(\omega(k'),0,\fs,J)v_1)\geq 0$.
Therefore,
$(y_a,y_{a+1})\leq (|y_a|/\sqrt{9}) |y_{a+1}|$ , so
\be
\label{sortho}
(y_a,y_{a+1})\leq (1/3) |y_a| |y_{a+1}|.
\ee
We define ${\cal W}$ to be the space spanned by all such vectors $y_a$,
and we define $P$ to project onto ${\cal W}$.
Consider any vector $v \in {\cal W}$, with
\be
v=\sum_{a=1}^{n_{seq}} v_a,
\ee
with $v_a$ parallel to $y_a$.
By Eq.~(\ref{sortho})
\be
\label{eisom2}
|v|^2 \geq 
\frac{1}{3} \sum_{a=1}^{n_{seq}} |v_a|^2.
\ee

{\bf Remark:} The function
$\ffn((\omega(i)+\omega(j))/2,
(\omega(j)-\omega(i))/2,\fs,\omega)$ is equal to unity for $\omega(i)\leq \omega \leq
\omega(j)$.

We now prove the Lemma (\ref{tridlemma})
as follows: to prove the first claim, note that by construction,
\begin{eqnarray}
|Pv_1-v_1|^2 
& \leq & |\sum_{a=1}^{n_{seq}} y_a -v_1|^2 \\ \nonumber
& \leq & 2 n_{win} \lambda_{min} \\ \nonumber
& \leq & 2/L^2.
\end{eqnarray}
The second line of the above equation follows because the difference
$\sum_{a} y_a-v_1$ is equal to $-\sum_{i {\rm unmarked}} 
\ffn(\omega(i),0,\fs,J)v_1$, where the sum ranges over
$i$ such that the corresponding window is unmarked.

To prove the second claim, consider the $a$-th sequence of
marked windows, from window $i$ to window $j$.  Let
$\omega_a=(\omega^-(i)+\omega^+(j))/2$.  Then, 
\be
|(J -\omega_a)y_a| \leq 
\Bigl(\frac{2+\lceil \log_{10/9}(2/\lambda_{min})\rceil}{n_{win}}\Bigr) |y_a|
\ee
which is bounded by $1/L$ times a function growing slower than any power of $L$.
Therefore,
\be
|(1-P)Jy_a| \leq
\Bigl(\frac{2+\lceil\log_{10/9}(2/\lambda_{min})\rceil}{n_{win}}\Bigr) |y_a|
\ee
Using the bound Eq.~(\ref{eisom2}),
for any vector $v \in {\cal W}$,
\be
|(1-P)Jv| \leq
2 \sqrt{3}
\Bigl(\frac{2+\lceil\log_{10/9}(2/\lambda_{min})\rceil}{n_{win}}\Bigr) |v|,
\ee
which is bounded by $1/L$ times a function growing slower than any power of $L$,
verifying the second claim.

The proof of the third claim is identical to the previous case.

\section{Quantum Measurement}
\subsection{Construction and Results}
The constructions above can be applied to operators which arise
in various physical quantum systems.  For example, consider a quantum
spin for a large spin $S$.  Then, the operators $S_x/S$ and $S_y/S$ have
operator norm 1 and have a commutator that is of order $1/S$.  Thus,
we can find a basis in which both operators are almost diagonal.  
While it is well known that one can use a POVM (positive operator-valued measure) to
approximately measure $S_x$ and $S_y$ at the same time,
the existence of the given basis implies that one can approximately
measure $S_x$ and $S_y$ simultaneously with a single {\it projective}
measurement.  
Interestingly, while the operator $S_z^2$ is also almost diagonal in this basis
(since it equals $S(S+1)-S_x^2-S_y^2$), it is not possible to find
a basis in which $S_x,S_y$, and $S_z$ are all almost diagonal (this
obstruction is similar
to that in \cite{voi}).  
Therefore, to approximately measure $S_x, S_y$, and $S_z$ simultaneously will
require a POVM, rather than a projective measurement.

For completeness, we now briefly show how to construct a POVM to approximately
measure several almost commuting operators simultaneously.
Consider any number $N$ of Hermitian matrices,
labeled
$A_1,...,A_N$, with $\Vert [A_i,A_j] \Vert \leq \delta$ for all $i,j$ and with
$\Vert A_i \Vert \leq 1$ for all $i$.  We now construct a POVM to approximately
measure all $N$ operators simultaneously.
The physical idea is very simple: we first do a ``soft" measurement of $A_N$,
then $A_{N-1}$, and so on, until all operators are measured.

Let $n_{win}$ be some integer given by
\be
n_{win}=\lceil \delta^{-1/2} (N-1)^{-1/2} \rceil
\ee
($n_{win}$ will typically be much
larger than unity).
For $i=1,...,N$ and $n=0,...,n_{win}-1$, define
\begin{eqnarray}
\omega(i)=-1+2i/(n_{win}-1)
&=& -1+i\fs,
\end{eqnarray}
where $\fs=2/(n_{win}-1)$
as before,
and define
\be
M(i,n)=\sqrt{\ffn(\omega(n),0,\fs,A_i)}.
\ee
The definition of $\ffn$ is given at the start of section IV; we will
see later that in this section that we do not actually need $\ffn$ to be
infinitely differentiable as it is defined there,
but we have only weaker requirements on $\ffn$.
Define 
\be
O(n_1,n_2,...,n_N)=\Bigl (M(1,n_1)^{\dagger} M(2,n_2)^{\dagger} ... M(N,n_N)^{\dagger}\Bigr)
\Bigl (M(N,n_N) ... M(2,n_2) M(1,n_1)\Bigr).
\ee
Then,
\be
\sum_{n_1,n_2,...=0}^{n_{win}-1} O(n_1,n_2,...,n_N)=\openone,
\ee
and all of the operators $O(n_1,n_2,...,n_N)$
are positive semidefinite by construction.  Therefore,
the operators $O(n_1,n_2,...,n_N)$ form a POVM.  Note that $M(i,n_i)=M(i,n_i)^{\dagger}$,
but we continue to write daggers on the operators for clarity.

We claim that this POVM approximately measures all operators simultaneously.
That is, we will show that for any density
matrix $\rho$, if the outcome of the measurement is $n_1,n_2,...,n_N$, then
if we perform
a subsequent measurement of any operator $A_i$, the outcome will be close to
$\omega(n_i)$ with high probability.  We show this by computing the
expectation value $(A_i-\omega(n_i))^2$ averaged over all measurement outcomes.
For any density matrix $\rho$, for any $i$, the average
over all outcomes of $(A_i-\omega(n_i))^2$ is equal to
\be
\label{avout}
\sum_{n_1,n_2,...=0}^{n_{win}-1} {\rm tr}\Bigl((A_i-\omega(n_i))^2 M(1,n_1) M(2,n_2) ...
\rho ... M(2,n_2)^{\dagger} M(1,n_1)^{\dagger}\Bigr)
\ee
The main result in this section is that
\be
\label{upperbound}
\sum_{n_1,n_2,...=0}^{n_{win}-1} {\rm tr}\Bigl((A_i-\omega(n_i))^2 M(1,n_1) M(2,n_2) ...
\rho ... M(2,n_2)^{\dagger} M(1,n_1)^{\dagger}\Bigr)\leq
{\rm const.} \times (N-1)\delta.
\ee
We show this in the next subsection.

\subsection{Bounds}

Note that $\Vert \sum_{n_i} M(i,n_i) \Vert \leq \sqrt{2}$.  
To bound Eq.~(\ref{avout}), we need three results,
Eqs.~(\ref{small},\ref{comm},\ref{comm1}) below.
First,
\begin{eqnarray}
\label{small}
\sum_{n_i=0}^{n_{win}-1} \Vert (A_i-\omega(n_i)) M(i,n_i) \Vert
& \leq &
{\rm const.} \times \fs
\\ \nonumber
& \leq &
{\rm const.} \times 1/n_{win}.
\end{eqnarray}

Second, we need
\begin{eqnarray}
\label{comm}
\Vert \sum_{n_j=0}^{n_{win}-1} 
[M(j,n_j),(A_i-\omega(n_i))] O 
M(j,n_j)^{\dagger}
\Vert
\leq {\rm const.} \times (\delta/\fs) \Vert O \Vert
\end{eqnarray}
for any operator $O$.

Third, we need
\begin{eqnarray}
\label{comm1}
\Vert \sum_{n_j=0}^{n_{win}-1} 
[M(j,n_j),(A_i-\omega(n_i))] 
O 
[M(j,n_j)^{\dagger},(A_i-\omega(n_i))]
\Vert
\leq {\rm const.} \times (\delta/\fs)^2 \Vert O \Vert
\end{eqnarray}
for any operator $O$.

Eq.~(\ref{small}) follows immediately from the support of $\ffn$.
To show Eq.~(\ref{comm}), define
\be
A^0=\fs \int {\rm d} t \exp(i A_j t) (A_i-\omega(n_i)) \exp(-i A_j t) f(\fs t),
\ee
where the function $f(t)$ is defined to have the Fourier transform
as in Eq.~(\ref{tfft}).
Then, $\Vert A^0-(A_i-\omega(n_i)) \Vert \leq {\rm const.} \times \delta/\fs$ as
in lemma (\ref{h0con}).  Also, if $v_1,v_2$ are eigenvectors of $A_j$ with
corresponding eigenvalues $x_1,x_2$ with $|x_1-x_2|\geq \fs$, then $(v_1,A^0 v_2)=0$,
which implies that
\be
\label{summit}
\Vert \sum_{n_j=0}^{n_{win}-1} 
[M(j,n_j),A^0] O 
M(j,n_j)^{\dagger}
\Vert
\leq 
2\, {\rm max}_{n_j} (\Vert 
[M(j,n_j),A^0] O 
M(j,n_j)^{\dagger}
\Vert).
\ee
Eq.~(\ref{summit}) is the reason for introducing the operator $A^0$.
We can bound the commutator
$[M(j,n_j),A^0]$ as follows.
Note that $\Vert [A_j,A^0] \Vert \leq \delta$.
Write 
\be
M(j,n_j)=\int{\rm d}t \exp(i A_j t)
\widetilde{\sqrt{\ffn(\omega(n_j),0,\fs,t)}},
\ee
where
$\widetilde{\sqrt{\ffn(.,.,.,t)}}$ denotes the Fourier transform of the
square-root of $\ffn$.
Then
since $\Vert [\exp(i A_j t),A^0]\Vert \leq
{\rm const.} \times |t| \delta$, we can use
a triangle inequality to show that
\be
\label{theint}
\Vert[M(j,n_j),A^0]\Vert
\leq \int {\rm d}t\, 
\widetilde{\sqrt{\ffn(\omega(n_j),0,\fs,t)}} |t| \delta.
\ee
Then, since $\sqrt\ffn(.,.,.,\omega)$ is infinitely differentiable,
the Fourier transform decays faster than any power of $t$ and
the integral over $t$ converges, so
we have
$\Vert [M(j,n_j),A^0] \Vert
\leq {\rm const.} \times \delta/\fs$.
Using Eq.~(\ref{summit}) gives Eq.~(\ref{comm}).
Eq.~(\ref{comm1}) is derived similarly.

Using Eqs.~(\ref{small},\ref{comm},\ref{comm1}), 
we can bound the sum in Eq.~(\ref{avout}) by writing $(A_i-\omega(n_i))^2=(A_i-\omega(n_i))(A_i-\omega(n_i))$, and commuting one of the terms $(A_i-\omega(n_i))$ to the right through
$M(j,n_j)$ for $j<i$ until it hits the $M(i,n_i)$ and commuting the
other term $(A_i-\omega(n_i))$ to the left through
$M(j,n_j)^{\dagger}$ for $j<i$ until it hits $M(i,n_i)^{\dagger}$.
Therefore,
\begin{eqnarray}
\label{cth}
&&\sum_{n_1,n_2,...=0}^{n_{win}-1} {\rm tr}((A_i-\omega(n_i))^2 M(1,n_1) M(2,n_2) ...
\rho ... M(2,n_2)^{\dagger} M(1,n_1))^{\dagger}
\\ \nonumber
& \leq &{\rm const}. \times \Bigl( (i-1)^2\delta^2 n_{win}^2+(i-1)\delta n_{win}/n_{win}+
1/n_{win}^2\Bigr)
\\ \nonumber
& \leq &{\rm const}. \times \Bigl( (N-1)^2\delta^2 n_{win}^2+(N-1)\delta+
1/n_{win}^2\Bigr).
\end{eqnarray}
The first term on the right-hand side of Eq.~(\ref{cth})
arises from two non-vanishing commutators (if the non-vanishing commutators are with $M(j,n_j)$ and
$M(k,n_k)^{\dagger}$ for $j\neq k$ then we use Eq.~(\ref{comm}) twice,
but if $j=k$ we use Eq.~(\ref{comm1}) once).
The second term arises from one non-vanishing commutator and one use of Eq.~(\ref{small}),
and the last term arises from using Eq.~(\ref{small}) twice.
Choosing
\be
n_{win}=\lceil \delta^{-1/2} (N-1)^{-1/2}\rceil,
\ee
we find that we measure all operators to within a mean-square error of
order $(N-1)\delta$, as claimed.

Note that we did not actually require that $\ffn(.,.,.,\omega)$ be infinitely
differentiable in this section.  We only required that the Fourier
transform $\widetilde{\sqrt{\ffn(.,.,.,t)}}$ decay sufficiently rapidly
in $t$ that the integral (\ref{theint}) converges.  The other properties of
$\ffn$ we used are that
$\sum_n \ffn(\omega(n),0,\fs,\omega)=1$ for $-1\leq \omega \leq 1$ and
that $\ffn(\omega(n),0,\fs,\omega)$ vanish for $|\omega-\omega(n)|\geq \fs$.

\section{Discussion}

The main result is a  polynomial relation between $\epsilon$ and $\delta$.
We have in fact implemented the
construction in Lemma (\ref{tridlemma})
for the uniform chain.  In practical applications,
we expect that, for many tridiagonal
matrices, the lack of orthogonality of the ${\cal X}_i$ will not
cause a problem, and choosing ${\cal W}$ to be the space spanned by
the ${\cal X}_i$ will lead to satisfactory results, without having to follow the
more complicated procedure above.  If, for some particular $J$,
the lack of orthogonality of the ${\cal X}_i$ does cause a problem,
an alternative procedure that might be more useful in practice
than the deterministic procedure above
is to add small, randomly chosen matrices to each diagonal
block of $J$.
This may smooth out the spectrum of $J$ and then allow one to
choose ${\cal W}$ to be the space spanned by the ${\cal X}_i$.

We gave above applications to quantum measurement.
Another application of this result is to construct Wannier functions
for any two dimensional quantum system for a spectral gap.  In
\cite{topo}, it was pointed out that given a two dimensional
quantum system with a gap between bands, one could define an operator
$G$ which projected onto the bands below the gap.  Then, define
the operator $X$ and $Y$ to measure $X$ and $Y$ position of particles,
and define $GXG$ and $GYG$ as projections of $X$ and $Y$ into the
lowest band.  Let $\Vert X \Vert,\Vert Y \Vert=L$, where
$L$ is the linear size of the system.
Since the operator $G$ was constructed in \cite{topo}
as a short-range operator, the commutator $\Vert [GXG,GYG] \Vert$
is small compared to $L^2$, and thus we can use the results here
to construct a basis of Wannier functions which is localized in both
the x- and y-directions.

{\it Acknowledgments---} I thank N. Filonov and I. Kachkovskiy for pointing out a mistake in the previous version.
I thank Y. Ogata for pointing out a gap in the proof on the previous
draft in the proof of the ``second property" (now the first property) and for many very useful discussions; while filling this gap in,
I found the modification of the construction here which in fact leads to tighter bounds.
I thank G. Bouch for useful comments.
I thank T. J. Osborne and J. Yard for
useful conversations, and I thank T. J. Osborne for
many comments on a draft of this paper.
I thank D. Poulin for explaining Jordan's lemma.
This work was supported by U. S. DOE Contract No. DE-AC52-06NA25396.


\begin{thebibliography}{99}
\bibitem{jl} Jordan's lemma states that given any two projectors, $M$ and $N$, we
can find a basis in which both projectors are block diagonal with blocks of
size $1$ or $2$.  Equivalently (this is the formulation we use), if $M$ projects onto
a subspace ${\cal M}$, then we can find an orthonormal basis of vectors $m_i$ for
${\cal M}$ with the property that $(m_i,N m_j)=0$ for $i \neq j$.  The original
reference is
C. Jordan: Essai sur la g\'eom\'etrie \`a $n$ dimensions.
Bulletin de la S. M. F. {\bf 3}, 103 (1875).
\end{thebibliography}
\end{document}